\documentclass[
 reprint,
 secnumarabic,twocolumn,
 amssymb,amsmath,nobibnotes,aps,prd,showkeys,showpacs,nofootinbib]{revtex4-2}
\usepackage{amsmath}
\usepackage{amssymb}
\usepackage{graphicx}
\usepackage{mathtools}
\usepackage{subfig}
\usepackage{lipsum} 
\usepackage[a-2u]{pdfx}

\usepackage{color}
\definecolor{darkgreen}{rgb}{0.0, 0.65, 0.31}
\usepackage[normalem]{ulem}

\begin{document}

\title{Action-angle formalism for extreme mass ratio inspirals in Kerr spacetime}

\author{Morteza Kerachian$^1$}
\email{kerachian.morteza@gmail.com}
\author{Luk\'{a}\v{s} Polcar$^{1,2}$}
\author{Viktor Skoup\'{y}$^{1,2}$}
\author{Christos Efthymiopoulos$^3$}
\author{Georgios Lukes-Gerakopoulos$^1$}
\email{gglukes@gmail.com}
\affiliation{${}^1$ Astronomical Institute of the Czech Academy of Sciences, Bo\v{c}n\'{i} II 1401/1a, CZ-141 00 Prague, Czech Republic}   
\affiliation{${}^2$ Institute of Theoretical Physics, Faculty of Mathematics and Physics, Charles University in Prague, 18000 Prague, Czech Republic}
\affiliation{${}^3$ Dipartimento di Matematica Tullio Levi-Civita, Universit\`{a} degli Studi di Padova, Via Trieste 63 35121 Padova, Italy}

\begin{abstract}
  We introduce an action-angle formalism for bounded geodesic motion in Kerr black hole spacetime using canonical perturbation theory. Namely, we employ a Lie series technique to produce a series of canonical transformations on a Hamiltonian function describing geodesic motion in Kerr background written in Boyer-Lindquist coordinates to a Hamiltonian system written in action-angle variables. This technique allows us to produce a closed-form invertible relation between the Boyer-Lindquist variables and the action-angle ones, while it generates in analytical closed form all the characteristic functions of the system as well. The expressed in the action-angle variable Hamiltonian system is employed to model an extreme mass ratio inspiral (EMRI), i.e. a binary system where a stellar compact object inspirals into a supermassive black hole due to gravitational radiation reaction. We consider the adiabatic evolution of an EMRI, for which the energy and angular momentum fluxes are computed by solving the Teukolsky equation in the frequency domain. To achieve this a new Teukolsky equation solver code was developed.
\end{abstract}

\maketitle
\section{introduction}\label{sec:intro}

An extreme mass ratio inspiral (EMRI) is a binary compact object system consisting of a secondary stellar body of mass $\mu$ inspiraling into a primary supermassive black hole of mass $M$, due to radiation reaction. The mass ratio $q=\mu/M$ of an EMRI lies in the range $10^{-7}\leq q \leq 10^{-4}$. EMRIs are one of the prominent sources for the planned space-based gravitational wave (GW) detectors like the Laser Interferometer Space Antenna (LISA) \cite{EMRIsLISA}. It is expected that we will be able to observe  $\sim 10^{4}-10^{5}$ cycles in the sensitivity frequency range. This implies that for modelling an EMRI the accuracy during one cycle must be below the $10^{-5}$ order \cite{drasco2005,EMRIsLISA}. 

The extreme difference in the mass ratio between the secondary and the primary allows us to treat the influence of the secondary as a perturbation to the primary's background spacetime \cite{Barack19,Pound22}. This background is expected to be sufficiently well described by a Kerr spacetime. Hence, perturbation theory in Kerr spacetime provides the means to model the radiation reaction driving an EMRI evolution.  

The modelling of the evolution can be split into two timescales \cite{Flanagan}: a slow one and a fast one. The slow timescale concerns the dissipation of energy and angular momentum due to the emission of gravitational waves, while the fast one concerns the orbital revolution of the secondary around the primary. In terms of celestial mechanics \cite{arnold2007mathematical}, the slow timescale deals with the actions of the system, while the fast one with the angles of the system. Realizing this one is compelled to express an EMRI system in action-angle (AA) variables. 

Schmidt in his seminal work called "Celestial Mechanics in Kerr spacetime" \cite{Schmidt02} was able to provide the fundamental frequencies of bounded geodesic motion in Kerr by using elements of the action-angle formalism. This in turn allowed the frequency domain decomposition of the Teukolsky equation \cite{Drasco04} providing the energy and angular momentum fluxes. The idea of using action-angle variables for EMRIs is now widely adapted \cite{Flanagan,Maarten14,LeTiec:2011ab,LeTiec:2015kgg,Fujita:2016igj,Isoyama:2018sib}, and there are works \cite{Fujita:2009bp, vandeMeent:2019cam} providing transformations from and to action-angle variables for bounded geodesic motion in Kerr spacetime using integrals and special functions,  but not in closed forms. Examples of how this non-closed form transformation can be implemented can be found in Ref.~\cite{Gair:2011,vandeMeent:2018,Lynch:2022}. Closed-form transformations have been given in the Schwarzschild spacetime for the first time in \cite{Witzany22} and \cite{Polcar22} using different approaches. Ref.~\cite{Witzany22} used a Taylor series approach to provide a closed-form transformation to and from action-angle variables, while Ref.~\cite{Polcar22} used a technique coming from  the canonical perturbation theory \cite{arnold2007mathematical,Efthymiopoulos11} to achieve this. In particular, Ref.~\cite{Polcar22} employed the Lie series approach, which we follow also in this work in order to provide a closed-form transformation to and from action-angle variables for bound geodesic orbits in Kerr spacetime.

Perturbation theory has enjoyed many successes spanning from celestial mechanics \cite{Morbidelli02}, like the prediction of Neptune's existence by Urbain Le Verrier in the mid nineteenth century, to general relativity \cite{Blanchet14,Barack19}. In our work, we employ two branches of the perturbation theory: the canonical perturbation theory renowned for the 
Kolmogorov-Arnold-Moser theorem \cite{Arnold63}, which allows us to express the Hamiltonian system describing the geodesic motion in Kerr spacetime purely in action terms; and the black hole perturbation one provided by Teukolsky in \cite{Teukolsky:1973ha}, which provides the action's fluxes due to gravitational radiation reaction. Combining these two perturbation theories we suggest in this work a scheme for adiabatic modelling of EMRIs aiming to contribute in the ongoing effort for more and more computationally efficient schemes \cite{Osburn2016,Katz21,Hughes21}. The framework that we setup in this work could be expanded to other directions by adding more perturbative sources \cite{Gair:2011}, like matter distribution around a black hole \cite{Polcar22} or the extended body effects induced by the secondary body \cite{Skoupy22}. Actually, the latter is expected to be the main advantage of the Lie series approach over other techniques providing action-angle variables for the motion in a Kerr background, see, e.g.. Ref.~\cite{Polcar22} for the Schwarzschild case.

The rest of the article is organized as follows. Sec.~\ref{sec:aafrmalizm} briefs basic concepts and steps of the Lie series canonical perturbation approach. The application of this approach on the Kerr geodesic motion is discussed in Sec.~\ref{sec:setup} and an approximate Hamiltonian system is introduced in action-angle variables. Sec.~\ref{sec:inspiral} provides adiabatic EMRI models using the introduced system in action-angle variables. Finally, Sec.~\ref{sec:Conc} summarizes our work.

\section{Action-angle formalism and Canonical perturbation theory}

We use in this work AA formalism and two elements from the canonical perturbation theory: the Lie series mapping and the Birkhoff normal form. This section briefly introduces these elements and for more on them, the interested reader is referred to the literature \cite {Flanagan,Efthymiopoulos11}. 

\subsection{Action-angle formalism}\label{sec:aafrmalizm}

For a Hamiltonian system which consists of a $2N$-dimensional differentiable manifold $\mathcal{M}$, there exists a symplectic coordinate system $\lbrace q_n, p_n\rbrace$ with $n=\lbrace 1, N\rbrace$. This Hamiltonian system $H(\boldsymbol{q}, \boldsymbol{p})$ is called an integrable system in an open region $\mathcal{U}$, if there exist $N$ integrals of motion which are independent and in involution at every point of $\mathcal{U}$. If the integrable Hamiltonian system is, moreover, compact and connected, then there should be a canonical transformation allowing us to write the system in \textit{action-angle} (AA) variables $\lbrace \psi_n, J_n\rbrace$. For the angles, it holds that
\begin{align}
    \psi_m+ 2 \pi \equiv \psi_m, \qquad 1\leq m \leq \kappa,
\end{align}
implying that either all ($\kappa=N$) or some ($\kappa<N$) of the angle variables are periodic~\cite{arnold2007mathematical,Fiorani}; the $J_m$ are the integrals of motion given by 
\begin{equation}\label{eq:defaction}
    J_i= \frac{1}{2 \pi} \oint p_i dq_i,
\end{equation}
where the path integral takes place over an irreducible circle. In AA variables an integrable system should read $H_{AA}(\boldsymbol{J})$, hence the Hamilton equations would be
\begin{equation}\label{eq:eqofmotiongen}
    \dot{\psi_i}=\frac{\partial H_{AA}(\boldsymbol{J})}{\partial J_i}\equiv \Upsilon_i(\boldsymbol{J}), \quad \dot{J_i}=-\frac{\partial H_{AA}(\boldsymbol{J})}{\partial \psi_i}=0.
\end{equation}
The solution to the above equations of motion is
\begin{align}\label{eq:soleqmotion}
    \psi_i=& \Upsilon_i \lambda+ \psi_{i0},\\
    J_i=& J_{i0},
\end{align}
where $J_{i0}$ and $q_{i0}$ are initial conditions and $\lambda$ is the evolution parameter.

In order to express  $H(\boldsymbol{q}, \boldsymbol{p})$ in the AA variables, i.e. $H_{AA}(\boldsymbol{J})$, we need to know the canonical transformation from the $\lbrace q_n, p_n\rbrace$ coordinate system to the $\lbrace \psi_m, J_m\rbrace$ one. However, determining the exact canonical transformation is not an easy task for most of the Hamiltonian systems. Thus, we use the canonical perturbation theory to approximate the system by using a series of canonical transformations. 

 \subsection{Lie Series}

If $\chi$ is a generating function, a Lie derivative for a given function $f$ is defined as 
\begin{align}
    f \rightarrow \mathcal{L}_\chi f= \lbrace f, \chi \rbrace,
\end{align}
where $\lbrace f, \chi \rbrace$ is the Poisson bracket, i.e. 
\begin{equation}
    \lbrace f, \chi \rbrace = \sum^{N}_{i=1} \left(\frac{\partial f}{\partial q_i}\frac{\partial \chi}{\partial p_i}-\frac{\partial \chi}{\partial q_i}\frac{\partial f }{\partial p_i} \right).
\end{equation}
The respective Lie series is defined as
\begin{equation}\label{eq:lieseries}
    \text{exp}(\mathcal{L}_{\chi}) f= \sum^{\infty}_{k=0} \frac{1}{k !} \mathcal{L}^k_{\chi} f.
\end{equation}
The Lie series transformation is a canonical one. 

\subsection{Birkhoff normal form theorem}

To derive a Hamiltonian in the action form, we follow the Birkhoff normal form theorem. Here we explain the necessary steps that one has to follow to derive the Hamiltonian in the action up to the desired order.

Let's assume that we have a Hamiltonian in the form
\begin{equation}\label{eq:hambkh}
    H^{(0)} = Z_0 (J_{i}^{0})+ \sum_{k=1}^N \epsilon^k H^{(0)}_{k}(\psi_{i}^{0},J_{i}^{0}),
\end{equation}
where the $\epsilon$ is called the book-keeping parameter and is used to keep track of the smallness of each term in the Hamiltonian. When all the calculations are done, the numerical value of the book-keeping parameter is set $\epsilon=1$. Note that $J_{i}^{0}$ are not constants of motion for the whole Hamiltonian~\eqref{eq:hambkh}, but just for the $Z_0$ part.

By expanding the Hamiltonian~\eqref{eq:hambkh} up to second order in $\epsilon$ we arrive at
\begin{align}\label{eq:hambkhexp}
    H^{(0)}&= Z_0(J_{i}^{0})+ \epsilon \left(Z_{1}^{(0)}(J_{i}^{0}) + h_{1}^{(0)}(\psi_{i}^{0},J_{i}^{0})\right)\nonumber\\
    +&\epsilon^2 \left(Z_{2}^{(0)}(J_{i}^{0}) + h_{2}^{(0)}(\psi_{i}^{0},J_{i}^{0})\right) + \mathcal{O}(\epsilon^3),
\end{align}
where each $H_k^{(0)}(\psi_{i}^{0},J_{i}^{0})$ was split to the pure action part $Z_{k}^{(0)}(J_{i}^{0})$ and the part $h_{k}^{(0)}(\psi_{i}^{0},J_{i}^{0})$, which still depends on angles, i.e.
\begin{equation}
    H_k^{(0)}(\psi_{i}^{0},J_{i}^{0})=Z_{k}^{(0)}(J_{i}^{0}) + h_{k}^{(0)}(\psi_{i}^{0},J_{i}^{0}).\nonumber
\end{equation}
For simplicity, we reduce the notation to $Z_{k}^{(0)},h_{k}$ dropping the respective dependencies. In the following, we provide an example of the procedure up to $\mathcal{O}( \epsilon^2)$.

To remove the angle dependency in the first order in $\epsilon$ in the Hamiltonian~\eqref{eq:hambkhexp}, we apply the Lie series as follows
\begin{align}\label{eq:H1}
    H^{(1)}=\text{exp}(\mathcal{L}_{\chi_1}) H^{(0)}= H^{(0)}+ \mathcal{L}_{\chi_{1}} H^{(0)}+\frac{1}{2} \mathcal{L}_{\chi_{1}}^2 H^{(0)},
\end{align}
where 
\begin{align}
    \mathcal{L}_{\chi_{1}} H^{(0)}&= \lbrace Z_0, \chi_1\rbrace + \epsilon \lbrace Z_{1}^{(0)} + h_{1}^{(0)}, \chi_1\rbrace\nonumber\\
    &+ \epsilon^2 \lbrace Z_{2}^{(0)} + h_{2}^{(0)}, \chi_1 \rbrace,\nonumber\\
    \mathcal{L}_{\chi_{1}}^2 H^{(0)}&= \lbrace \lbrace Z_0, \chi_1\rbrace, \chi_1 \rbrace + \epsilon\lbrace \lbrace Z_{1}^{(0)} + h_{1}^{(0)}, \chi_1\rbrace,\chi_1\rbrace\nonumber\\
    &+ \epsilon^2 \lbrace\lbrace Z_{2}^{(0)} + h_{2}^{(0)}, \chi_1 \rbrace,\chi_1\rbrace.
\end{align}
We choose $\chi_1$ to be a quantity of the order $\mathcal{O}(\epsilon)$; therefore,
\begin{align}
    Z_0&= \mathcal{O}(0),\\
    \epsilon& \left(Z_{1}^{(0)} + h_{1}^{(0)}\right)+ \lbrace Z_0,\chi_1\rbrace = \mathcal{O}(\epsilon),\label{eq:eps1}\\
    \epsilon^2& \left(Z_{2}^{(0)} + h_{2}^{(0)}\right)+ \epsilon \lbrace Z_{1}^{(0)} + h_{1}^{(0)}, \chi_1\rbrace\nonumber\\
    +&\frac{1}{2} \lbrace \lbrace Z_0, \chi_1\rbrace, \chi_1 \rbrace = \mathcal{O}(\epsilon^2),\label{eq:esp2}
\end{align}
and in a similar way for $\mathcal{O}(\epsilon^3)$. 

From Eq.~\eqref{eq:eps1} we see that to eliminate the angle dependency at order
$\mathcal{O}(\epsilon)$,  we have to define $\chi_1$ as 
\begin{equation}\label{eq:chi1}
    \epsilon h_1^{(0)}+ \lbrace Z_0,\chi_1\rbrace=0.
\end{equation}
This equation and the respective equations that eliminate the angle dependency for higher orders in $\epsilon$ are called homological equations. By substituting $\chi_1$ from~\eqref{eq:chi1} into the Eq.~\eqref{eq:H1}, we can rewrite the Hamiltonian~\eqref{eq:H1} in its normal form up to the $\mathcal{O}( \epsilon)$:
\begin{align}\label{eq:H1nf}
    H^{(1)} & = Z_0+ \epsilon Z_{1}^{(0)}+\epsilon^2 \left(Z_{2}^{(1)} + h_{2}^{(1)}\right)+ \mathcal{O}(\epsilon^3), 
\end{align}
where 
\begin{align}
\epsilon^2 & \left(Z_{2}^{(1)} + h_{2}^{(1)}\right)= \epsilon^2 \left(Z_{2}^{(0)} + h_{2}^{(0)}\right)\nonumber\\
+ &\epsilon \lbrace Z_{1}^{(0)} + h_{1}^{(0)}, \chi_1\rbrace+\frac{1}{2} \lbrace \lbrace Z_0, \chi_1\rbrace, \chi_1 \rbrace.
\end{align}
To bring the Hamiltonian~\eqref{eq:H1nf} normal form up to $\mathcal{O}( \epsilon^2)$, we have to apply another Lie series in a similar fashion as we did for $\mathcal{O}(\epsilon)$. We start from 
\begin{equation}\label{eq:H2}
    H^{(2)}=\text{exp}(\mathcal{L}_{\chi_2}) H^{(1)}= H^{(1)}+ \mathcal{L}_{\chi_{2}} H^{(1)}+\frac{1}{2} \mathcal{L}_{\chi_{2}}^2 H^{(1)},
\end{equation}
where 
\begin{align}
    \mathcal{L}_{\chi_{2}} H^{(1)}&= \lbrace Z_0, \chi_2\rbrace + \epsilon \lbrace Z_{1}^{(0)} , \chi_2\rbrace+ \epsilon^2 \lbrace Z_{2}^{(1)} + h_{2}^{(1)}, \chi_2 \rbrace,\nonumber\\
    \mathcal{L}_{\chi_{2}}^2 H^{(1)}&= \lbrace \lbrace Z_0, \chi_2\rbrace, \chi_2 \rbrace + \epsilon\lbrace \lbrace Z_{1}^{(0)}, \chi_2\rbrace,\chi_2\rbrace\nonumber\\
    &+ \epsilon^2 \lbrace\lbrace Z_{2}^{(1)} + h_{2}^{(1)}, \chi_2 \rbrace,\chi_2\rbrace.
\end{align}
By choosing $\chi_2$ to be a quantity of $\mathcal{O}(\epsilon^2)$, the generating function $\chi_2$ can be determined from the second homological equation
\begin{equation}\label{eq:chi2}
    \epsilon^2 h_2^{(1)}+ \lbrace Z_0,\chi_{2}\rbrace=0.
\end{equation}
Using the solution of Eq.~\eqref{eq:chi2} and apply it into the Hamiltonian~\eqref{eq:H2} we arrive at the normal form
\begin{equation}
    H^{(2)}= Z_0+\epsilon Z_1^{(0)}+ \epsilon^2 Z_2^{(1)}+ \mathcal{O}(\epsilon^3),
\end{equation}
The final step is to set $\epsilon=1$. The $ \mathcal{O}(\epsilon^3)$ part still depends on angles and it is called the remainder.

\subsection{General formalism and benefit}\label{sec:genralformalism}

In the previous section, we demonstrated in detail how to express a Hamiltonian in AA variables using the Lie canonical transformation up to the $\mathcal{O}(\epsilon^2)$. However, we can generalize this computation to an arbitrary order $H^{(n)}$ by applying $n$ canonical transformations. The $H^{(n)}$ then reads
\begin{align}\label{eq:Hr}
    H^{(n)}=\text{exp}(\mathcal{L}_{\chi_n})\,\text{exp}(\mathcal{L}_{\chi_{n-1}})...\text{exp}(\mathcal{L}_{\chi_2})\, \text{exp}(\mathcal{L}_{\chi_1}) H^{(0)}.
\end{align}
In Eq.~\eqref{eq:Hr} each $\chi_\kappa$ where $\kappa \in \lbrace 1, n \rbrace$ comes from a solution of the respective homological equation, i.e. 
\begin{equation}\label{eq:chikap}
    \epsilon^\kappa h_\kappa^{(\kappa-1)}+ \lbrace Z_0, \chi_\kappa\rbrace=0.
\end{equation}
The main benefit of using the Lie series is that we can always transform the original variables (sometimes we call them old variables) to the AA variables (new variables) and vice versa. The transformation from the old variables to the new ones for $n$ transformations is given by a Lie series composition in the following way
\begin{equation}\label{eq:oldtonew}
    X_{\text{new}}=\text{exp}(\mathcal{L}_{\chi_n})\,\text{exp}(\mathcal{L}_{\chi_{n-1}})...\text{exp}(\mathcal{L}_{\chi_1}) X_{\text{old}},
\end{equation}
and the transformation from new variables to old variables is given by
\begin{equation}\label{eq:newtoold}
    X_{\text{old}}=\text{exp}(-\mathcal{L}_{\chi_1})...\text{exp}(-\mathcal{L}_{\chi_{n-1}})\,\text{exp}(-\mathcal{L}_{\chi_n})  X_{\text{new}},
\end{equation}
where $X$ can be any phase space quantity like action, angle, coordinate,   or momentum.

\section{Expressing geodesic motion in Kerr in AA variables}\label{sec:setup}

 \begin{figure}
    \begin{center}
        \includegraphics[width=0.483\textwidth]{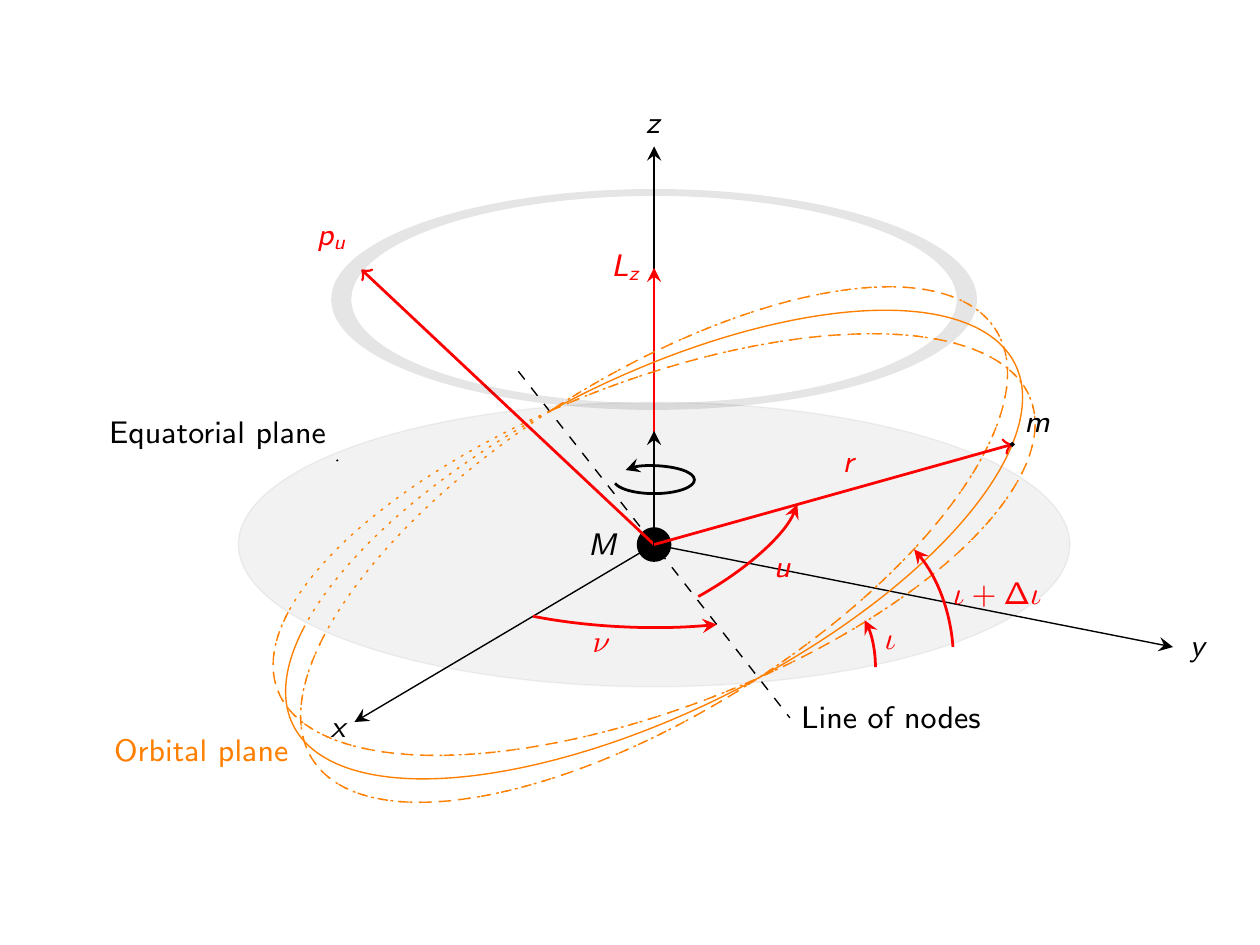}
    \caption{This figure shows an off-equatorial motion of a secondary object with mass $m$ around the Kerr black hole with mass $M$ in the polar nodal coordinates. The radial distance of the secondary from the supermassive black hole is indicated by $r$, and the angle $u$ represents the azimuth-like angle on the instantaneous orbital plane. Due to the spin of the black hole, the orbital plane itself precesses around the $z$-axis with the precession angle $\nu$ and its inclination $\iota$  oscillates slightly by $\Delta \iota$. The $z$ component of the orbital angular momentum is conserved, while the orbital angular momentum, whose measure is denoted as $p_u$ and which is perpendicular to the instantaneous orbital plane, is oscillating. The grey ring is the area that the $p_u$ is swiping while the secondary object rotates around the black hole. The line of nodes is the intersection of the orbital plane and the equatorial plane. }       \label{fig:polnod}
    \end{center}
\end{figure}

The  Kerr line element in Boyer-Lindquist coordinates $(t,r,\theta,\phi)$ reads
\begin{align}
        ds^2= -&\left( 1-\frac{2 M r}{\Sigma} \right) dt^2+ \frac{\Sigma}{\Delta}\,dr^2+ \Sigma\,d\theta^2 \nonumber\\
        +& \left( r^2+a^2+\frac{2 M a^2 r}{\Sigma} \sin^2 \theta \right) \sin^2\theta\, d\phi^2
       \nonumber \\-& \frac{4 M a r }{\Sigma} \sin^2\theta\, dt\,d\phi,
\end{align}
where $a$ is the Kerr parameter corresponding to the spin of the black hole and $M$ is its mass, while
\begin{align*}
    \Delta= r^2- 2 M r+a^2, \qquad \Sigma= r^2+a^2 \cos^2\theta.
\end{align*}
The stationarity and the axisymmetry of the spacetime provide two integrals of motion, the energy $E$ and the angular momentum $L_z$ along the symmetry axis $z$. The third integral is the Carter constant $Q$ \cite{Carter:1968rr}, which in the case of $ a=0$  is related to the  projection of the angular momentum on the equatorial plane ($Q\rvert_{a=0}=L_x^2+L_y^2$). The equations of the geodesic motion in a spacetime, which is described by a metric tensor $g_{\mu\nu}$,  can be provided by the Hamiltonian function
\begin{equation}
    H= \frac{1}{2} g^{\mu\,\nu} p_\mu p_\nu.
\end{equation}
For the Kerr spacetime, this Hamiltonian reads 
\begin{align}\label{eq:Hamiltonian}
    H=& \frac{\Delta}{2 \Sigma} p_{r}^2+\frac{1}{2\, \Sigma} p_{\theta}^2+\frac{(p_{\phi}+ a \, \sin^2 \theta p_t)^2}{2\, \Sigma\, \sin^2 \theta}\nonumber\\
    &-\frac{\left( \left( r^2+a^2\right)\, p_t+a \, p_{\phi}\right)^2}{2\,\Sigma \,\Delta},
\end{align}
since the system is autonomous the Hamiltonian itself is an integral of motion. Having four independent and in involution constants in a four degrees of freedom system implies that the system is integrable and, as Carter showed, separable \cite{Carter:1968rr}. This suggests that, in principle, there should be a way to express the Hamiltonian system in AA variables through a canonical transformation.

Actually, the Hamiltonian~\eqref{eq:Hamiltonian} can be separated into a radial and an angular part by replacing the proper time $\tau$ with the Carter-Mino \cite{Carter:1968rr,Mino} time $\lambda$, i.e. $d\tau = (r^2+a^2 \cos^2 \theta) d \lambda$. Then, the Hamiltonian~\eqref{eq:Hamiltonian} transforms to
\begin{multline}\label{eq:hammino}
        H_\lambda=\frac{1}{2}\left(\Delta p_r^2-\frac{((r^2+a^2)\,p_t+a L_z)^2}{\Delta}+\mu^2 r^2 \right)\\
        +\frac{1}{2} \left( p_\theta^2+a^2 \mu^2 \cos^2 \theta+\frac{(L_z+a \sin^2\theta p_t)^2}{\sin^2\theta}\right).
\end{multline}
The next step we take is to express Hamiltonian~\eqref{eq:hammino} in polar-nodal coordinates\footnote{For more on polar-nodal coordinates see Appendix~\ref{sec:polnod}}. This choice of coordinates provides a good insight into the motion of the secondary, since the coordinates and their conjugate momenta are related to the orbital parameters. For instance, in the case of the Schwarzschild spacetime $(a=0)$,  an orbit takes place on a single plane that has a fixed inclination angle $\iota$, a fixed precession angle $\nu$, and the orbital angular momentum of measure $p_u$ is perpendicular to this orbital plane. Moreover, in the polar-nodal coordinates representation, the motion is split into two parts: the motion of the secondary on the orbital plane described by the coordinates $r$ and $u$; and the motion of the orbital plane itself described by the angles $\nu$ and $\iota$; in this set up  $\lbrace t, r, u, \nu\rbrace$ are the coordinates and their conjugate momenta are $\lbrace p_t, p_r, p_u, L_z\rbrace$. Fig.~\ref{fig:polnod} illustrates these variables.

The coordinate change is provided by a canonical transformation given by
  \begin{align}\label{eq:cantrans}
      \theta&= \arccos \left ( \sqrt{1-\frac{L_z^2}{p_u^2}} \sin u \right),\nonumber\\
      p_{\theta}&= p_u \sqrt{1-\frac{L_z^2}{p_u^2-(p_u^2-L_z^2) \sin^2 u}},\\
    \phi&= u+\nu + \arctan\left(\frac{(L_z/p_u-1) \cos u \sin u}{1+ (L_z/p_u-1) \sin^2 u} \right),\nonumber
\end{align}
where $u$ is the azimuth-like angle in the inclined plane of the orbital motion. Note that the angular momentum $L_z$ remains unchanged in this canonical transformation. With the help of $p_u$ we can define the inclination angle as
 \begin{equation} \label{eq:inclinLzpu}
     \cos \iota = \frac{L_z}{p_u}.
 \end{equation}
Note that this definition is different than $\cos \iota = L_z/\sqrt{L_z^2+Q}$ given in \cite{Hughes00}; since $p_u$ is not constant in Kerr spacetime, the inclination angle oscillates. 

Consequently, the Hamiltonian in the polar-nodal coordinate will be 
\begin{align}\label{eq:hampolnod}
            &H_{pn}=\frac{1}{2}\left(\Delta\, p_r^2-\frac{((r^2+a^2)\,p_t+a L_z)^2}{\Delta}+\mu^2 \,r^2 \right)\\
          +&\dfrac{1}{2}\bigg( a ^2 p_t^2+ 2\,a\,L_z\,p_t+ p_u^2+ \frac{a^2 (\mu^2 -p_t^2)\,(p_u^2-L_z^2)\sin^2 u}{p_u^2}\bigg),\nonumber
\end{align}

We could apply the Lie series at this point, but we prefer to rewrite first the angular part in terms of the Carter constant as follows
\begin{align}\label{eq:hamcart}
            H_{pn}=&\frac{1}{2}\left(\Delta\, p_r^2-\frac{((r^2+a^2)\,p_t+a L_z)^2}{\Delta}+\mu^2\,r^2 \right)\nonumber\\
           +\frac{1}{2}& \left((a p_t+L_z)^2+ Q\right),
\end{align}
where
\begin{equation}\label{eq:carter}
    Q= \left(1-\frac{L_z^2}{p_u^2}\right)\,\left(p_u^2+a^2\,(\mu^2 -p_t^2)\,\sin^2 u \right)
\end{equation}
is the Carter constant in the polar-nodal coordinates. 

Since $H_{pn}=0$, we can get $p_r$ as a function of $r,~a,~\mu$ and the constants of motion, where $\mu$ is the mass of test body; from Eq.\eqref{eq:carter} $p_u$ can be written as a function of $u,~a,~\mu$ and the constants of motion. This leads to the following relations  
\begin{align}
    p_t= -E,\,\, p_r= \pm \frac{\sqrt{V_r}}{\Delta},\,\,p_u= \sqrt{\frac{V_u}{2}},\,\, p_\nu= L_z.
\end{align}
The $V_r$ and $V_u$ are
\begin{align}
    V_r=& [a\,L_z-(r^2+a^2) E]^2-\Delta [\mu^2\,r^2+(L_z-a\,E)^2+Q],\nonumber\\
   V_u=& v_u+\sqrt{4\,L_z^2\,(\mu^2-E^2)a^2\,\sin^2 u +v_u^2},
\end{align}
where $$   v_u= L_z^2+Q-(\mu^2-E^2)\,a^2\,\sin^2 u.$$

In polar-nodal coordinate, the four constants of motions are $\lbrace \overline{J}_r, \overline{J}_u, \overline{J}_{\nu}, \overline{J}_t\rbrace$ and their conjugate angle are denoted by $\lbrace \overline{\psi}_r, \overline{\psi}_u, \overline{\psi}_\nu, \overline{\psi}_t\rbrace$; here the over-line refers to the definition of the actions and the angles~\eqref{eq:defaction}-~\eqref{eq:soleqmotion}. These actions are defined from Eq.~\eqref{eq:defaction} as follows
\begin{align}
    \overline{J}_r&= \frac{1}{2 \pi} \oint \frac{\sqrt{V_r}}{\Delta} dr=\frac{1}{ \pi} \int_{r_p}^{r_a} \frac{\sqrt{V_r}}{\Delta} dr ,\label{eq:jrpl}\\
    \overline{J}_u&= \frac{1}{2 \pi}\oint \sqrt{\frac{V_u}{2}} du=\frac{1}{ \pi}\int_0^{\pi} \sqrt{\frac{V_u}{2}} du,\label{eq:jupl}\\
    \overline{J}_\nu&=\frac{1}{2 \pi}\oint p_\nu d\nu= L_z,\label{eq:jnupl}\\
    \overline{J}_t&=\frac{1}{2 \pi} \int_0^{2 \pi} p_t dt= -E\label{eq:jtpl},
\end{align}
where  $r_{p,a}$ are the periapsis and apoapsis, which are related to the semi-latus rectum $p$ and eccentricity $e$  as
\begin{align}
    r_{p,a}= \frac{p}{1\pm e}.
\end{align}
 We got numerical indications that $\overline{J}_u= \overline{J}_\nu+ \overline{J}_\theta $ , where $\overline{J}_\theta$ is the action related to the $\theta$ angle in the Boyer-Lindquist coordinates~\cite{Flanagan}. This implies that $\overline{J}_u$ is the action that reflects the total angular momentum. Note that in our work by assumption $L_z>0$.

Hamiltonian~\eqref{eq:hamcart} has been reduced to just one degree of freedom along the radius, since the angular dependence is hidden in the Carter constant. We take advantage of the above split into radial and angular parts in the following sections.

\subsection{Radial motion}

In order to apply the Lie series to the Hamiltonian~\eqref{eq:hamcart}, we should convert it to the form of Eq.~\eqref{eq:hambkh}. To do that, we first have to choose a reference orbit around which we shall expand the system perturbatively. We choose to expand around a spherical orbit in Kerr \cite{Hughes00}, i.e. an orbit of constant radius. Since the Hamiltonian~\eqref{eq:hamcart} corresponds to a system of one degree of freedom, a spherical orbit degenerates to a circular one. This orbit can be characterized by its radius $r_c$ and angular momentum $L_{zc}$, where the index "c" refers to a circular orbit. The Carter constant and the energy can be derived by the following relations:
\begin{widetext} 
\begin{align} 
 E_c&= -p_{tc}= \frac{a^{2}\, L_{zc}^2 \,(r_c-M)+ r_c\, \Delta_c^2}{a\, L_{zc}\,M\,(r_c^2-a^2) +\Delta_c \sqrt{r_c^5\,(r_c-3\,M)+a^4\,r_c\,(r_c+M)+a^2 r_c^2\,(L_{zc}^2-2\,M\,r_c+2\,r_c^2)}}, \label{eq:Ec}\\
 Q_c&= \frac{\left( (a^2+r_c^2) E_c-a\,L_{zc} \right)^2}{\Delta_c}-\left(r_c^2+a^2\,E_c^2-2\,a\,E_c\,L_{zc}+L_{zc}^2 \right), \label{eq:Qc}
\end{align}
\end{widetext}
where $\Delta_c= r_c^2- 2\,M\,r_c+a^2$. 

Now, from a fixed reference orbit, we parameterize the deviation from it as follows
\begin{align}\label{eq:bkkp}
        r&= r_c+\epsilon \, d\hat{r}, \qquad p_r= \epsilon\,\hat{p}_r,  \qquad L_z= L_{zc}+ \epsilon^2\, J_{\nu}^0,  \nonumber\\
        p_t&= p_{tc}+\epsilon^2\, J_t^0, \qquad Q=Q_{c}+\epsilon^2\, \Tilde{Q}. 
\end{align}
Note that we expand around $p_{t}$ instead of the energy. 

The accuracy of the scheme highly depends on the choice of the position of the circular orbit $r_c$ and how we set the small perturbation $d\hat{r}$. By trial and error, we have found that by using the canonical transformation
\begin{equation}\label{eq:accuracytrans}
    d\hat{r}= \delta dr, \qquad \hat{p}_r= \frac{\Tilde{p}_r}{\delta}.
\end{equation}
and appropriately choosing $r_c$ and $\delta$ we could improve the accuracy of the scheme. 

We substitute the transformations~\eqref{eq:bkkp}-~\eqref{eq:accuracytrans} into the Hamiltonian~\eqref{eq:hamcart} and expand it with respect to the book-keeping parameter $\epsilon$. This leads to
\begin{equation}\label{eq:hamser}
     H_o=\left(\Omega_{t0} J_t^0 + \frac{1}{2} \,\Tilde{Q}+\Omega_{z0} J_{\nu}^0 + \alpha\, dr^2 + \beta \,\Tilde{p}_r^2 \right) \epsilon^2 +\mathcal{O}(\epsilon^n), 
\end{equation}
where  $n\geq 3$. Note that the zero and linear terms in $\epsilon$ vanished. The explicit formulas of $\Omega_{t0}$, $\Omega_{z0}$, $\alpha$ and $\beta$ read
\begin{align}\label{eq:fisrtconst}
\Omega&_{t0}=-\frac{r_c}{\Delta_c}\big(p_{tc}(r_c^3+a^2(2 M+r_c))+2\,a\,M\,L_{zc} \big),\nonumber\\
\Omega&_{z0}= \frac{r_c}{\Delta_c}\big(L_{zc}(r_c-2\,M)-2\,a\,M\,p_{tc}\big), \,\beta=-\frac{\Delta_c}{2 \delta^2}\\
\alpha&= \frac{-\delta^2}{2\Delta_{c}}\Big(\mathfrak{A}_1\,p_{tc}^2+4\,\mathfrak{A}_2\,a\,M\,L_{zc}\,p_{tc}+\mathfrak{A}_3\,a^2\,L_{zc}^2-\Delta_c^3\Big),\nonumber
\end{align}
where
\begin{align}
    \mathfrak{A}_1&= \Delta_c^3+4\,M^2(a^4-3a^2r_c^2+2\,M\,r_c^3)\nonumber\\
\mathfrak{A}_2&=r_c^3+a^2(2 M-3 r_c),\nonumber\\
\mathfrak{A}_3&=a^2(M^2-a^2)+3 \Delta_c.\nonumber
\end{align}

Since in the part of the Hamiltonian~\eqref{eq:hamser} up to $\mathcal{O}(\epsilon^2)$ apart from the $(dr,~\Tilde{p}_r)$ pair all the other quantities  are constants, this part of the Hamiltonian is basically describing a harmonic oscillator, where the mass and frequency can be given respectively as
\begin{align}
    m_c= \frac{1}{2 \beta}, \qquad \Omega_{r0}= \sqrt{\frac{2 \alpha}{m_c}}. 
\end{align}
A harmonic oscillator can be expressed into AA variables by the well-known transformation
\begin{align}\label{eq:harmonictrans}
    dr&= \sqrt{\frac{2\, J_{r}^{0}}{m_c\, \Omega_{r0}} }\sin \psi_r^0,\\
    \Tilde{p}_r&= \sqrt{2 \,J_{r}^{0}\, m_c\, \Omega_{r0}} \cos \psi_r^0.
\end{align}
We substitute the above relations into Eq.~\eqref{eq:hamser} and, since the zero and linear terms have vanished in Eq.~\eqref{eq:hamser}, we reduce the order of the Book-keeping parameter by $2$. This results in
\begin{equation}\label{eq:hamser2}
     H^{(0)}=\left(\Omega_{t0}\, J_t^0 + \frac{1}{2} \,\Tilde{Q}+\Omega_{z0}\, J_{\nu}^0 + \Omega_{r0}\,J_{r}^{0}\right) +\mathcal{O}(\epsilon^n),   
\end{equation}
where $n\geq 1$. Note that the Hamiltonian~\eqref{eq:hamser2} is in the form of Eq.~\eqref{eq:hambkh}, since at order  $\epsilon^0$ the $\lbrace J_t^0, J_{\nu}^0, J_r^0\rbrace$ are the actions of the system. We denote their conjugate angles as $\lbrace \psi_t^0, \psi_\nu^0, \psi_r^0\rbrace$. Therefore, we can now apply the Lie series, derive the generating functions, and express the Hamiltonian~\eqref{eq:hamcart} in terms of AA variables. Note that $\Tilde{Q}$ is a constant, but not an action. We address this issue in the next section. 
 
\subsection{Angular motion}
 
In the previous section, we expanded the Hamiltonian~\eqref{eq:hamcart} around a spherical orbit in Kerr. In this section, we apply similar steps to express $\Tilde{Q}$ in terms of the actions. For the Carter constant, however, we choose a circular orbit in Schwarzschild as a reference orbit and we expand the system around it perturbatively. This orbit is characterized by its radius $r_{c_{\rm sc}}$, total angular momentum $p_{uc}$, z-component of the angular momentum $\Tilde{L}_{zc}$ and its energy $\Tilde{E}_c= - \Tilde{p}_{tc}$. By substituting $a=0$ into the Eq.~\eqref{eq:Ec} and Eq~\eqref{eq:Qc} we get $E_c= \Tilde{E}_c$ and $Q_{c}=\Tilde{Q}_{c}=p_{uc}^2-\Tilde{L}_{zc}^2$\footnote{Note that, for the expansion around the Schwarzschild circular orbit in the angular motion, i.e., around $r_{c_{\rm sc}}$, we use the tilde symbol of the quantities to distinguish them from the expansion around the Kerr circular orbit in the radial motion, i.e., around $r_{c}$.}. The deviation from this reference orbit is parameterized as follows: 
\begin{align}\label{eq:carbk}
 p_u&= p_{uc}+\epsilon^2\, J_{u}^0, \qquad {L}_{z}= \Tilde{L}_{zc}+\epsilon^2 \Tilde{J}_\nu^0, \nonumber\\
 {p}_t&= \Tilde{p}_{tc}+ \epsilon^2 \Tilde{J}_t^0, \qquad a= \epsilon^2 \Tilde{a}.
\end{align}

The Carter constant~\eqref{eq:carter} can be set as the Hamiltonian describing the angular motion. Having this in mind, we expand it with respect to $\epsilon$ to arrive at
\begin{equation}\label{eq:carbkser}
    Q=  (p_{uc}^2-\Tilde{L}_{zc}^2)+ 2 (p_{uc} \,J_{u}^0+ \Tilde{L}_{zc}\, \Tilde{J}_\nu^0) \,\epsilon^2+ \mathcal{O}(\epsilon^{n}),
\end{equation}
for $n\geq 3$. Since $p_{uc}^2-\Tilde{L}_{zc}^2$ is constant, we can remove it and define a new Hamiltonian $Q^{(0)}= Q- (p_{uc}^2-\Tilde{L}_{zc}^2)$.  Furthermore, we can reduce the order of the Book-keeping parameter $\epsilon$ by 2. Hence, we arrive at 
\begin{equation}\label{eq:carbkser2}
    Q^{(0)}=  2 (p_{uc} \,J_{u}^0+ \Tilde{L}_{zc}\, \Tilde{J}_\nu^0)+\mathcal{O}(\epsilon^{n}),
\end{equation}
where $n\geq 1$. Now, the Carter constant~\eqref{eq:carbkser2}  is in the form of~\eqref{eq:hambkh}, since at order $\epsilon^0$ the $\lbrace J_u^0, \Tilde{J}_{\nu}^0\rbrace$ are the actions of the system. Their conjugate angles are denoted by $\lbrace \psi_u^0, \Tilde{\psi}_\nu^0\rbrace$, while $\Tilde{\psi}_t^0$ is the conjugate angle to the action $\Tilde{J}_t^0$. At this point, we can apply the Lie series to the system. 

\subsection{New Hamiltonian} \label{sec:NewHam}

By applying $n$ canonical transformations, i.e. Lie series transformations, to the Hamiltonian~\eqref{eq:hamser2} and $n^\prime$ canonical transformations to the Carter constant~\eqref{eq:carbkser2}, the new Hamiltonian takes the form
\begin{equation} \label{eq:Haa}
    H_{AA}= \mathcal{H}(J_t,J_{\nu}, J_r) + \frac{1}{2} \Tilde{Q}+ \mathcal{O}(\epsilon^{n+1}),
\end{equation}
with 
\begin{equation}\label{eq:cartaa}
      \Tilde{Q}= \mathcal{Q}(J_u, \Tilde{J}_t, \Tilde{J}_\nu)+  (p_{uc}^2-\Tilde{L}_{zc}^2)- Q_c+\mathcal{O}(\epsilon^{ n^{\prime}+1}),
\end{equation}
where the $\lbrace J_r,J_u, J_{\nu}, J_t \rbrace$ are the actions of the new system\footnote{We substitute $\Tilde{J}_\nu= J_{\nu}+ L_{zc}- \Tilde{L}_{zc}$ and $\Tilde{J}_{t}= J_t+ p_{tc}-\Tilde{p}_{tc}$ into the Eq.~\eqref{eq:cartaa}. Therefore, the Hamiltonian~\eqref{eq:Haa} is in terms of the $\lbrace J_t, J_r,J_u, J_{\nu} \rbrace$. Note that, the conjugate angles for this system are denoted by $\lbrace \psi_r, \psi_u, \psi_\nu, \psi_t\rbrace$. }; the   $\mathcal{O}(\epsilon^{n+1})$ and $\mathcal{O}(\epsilon^{n^\prime+1})$ are the remainders. When we refer to the $H_{AA}$ Hamiltonian from now on we consider it without the remainders. At this point, we set the book-keeping parameter $\epsilon=1$ in the $H_{AA}$. The exact expression of Eq.~\eqref{eq:Haa} is extremely long to be provided in the text, thus we provide it in the supplemental material \cite{CPKerrGeodesics}.

\begin{figure*}[ht]
\begin{center}
\captionsetup[subfloat]{position=top} 
 {\subfloat[]{\includegraphics[width=0.48\textwidth]{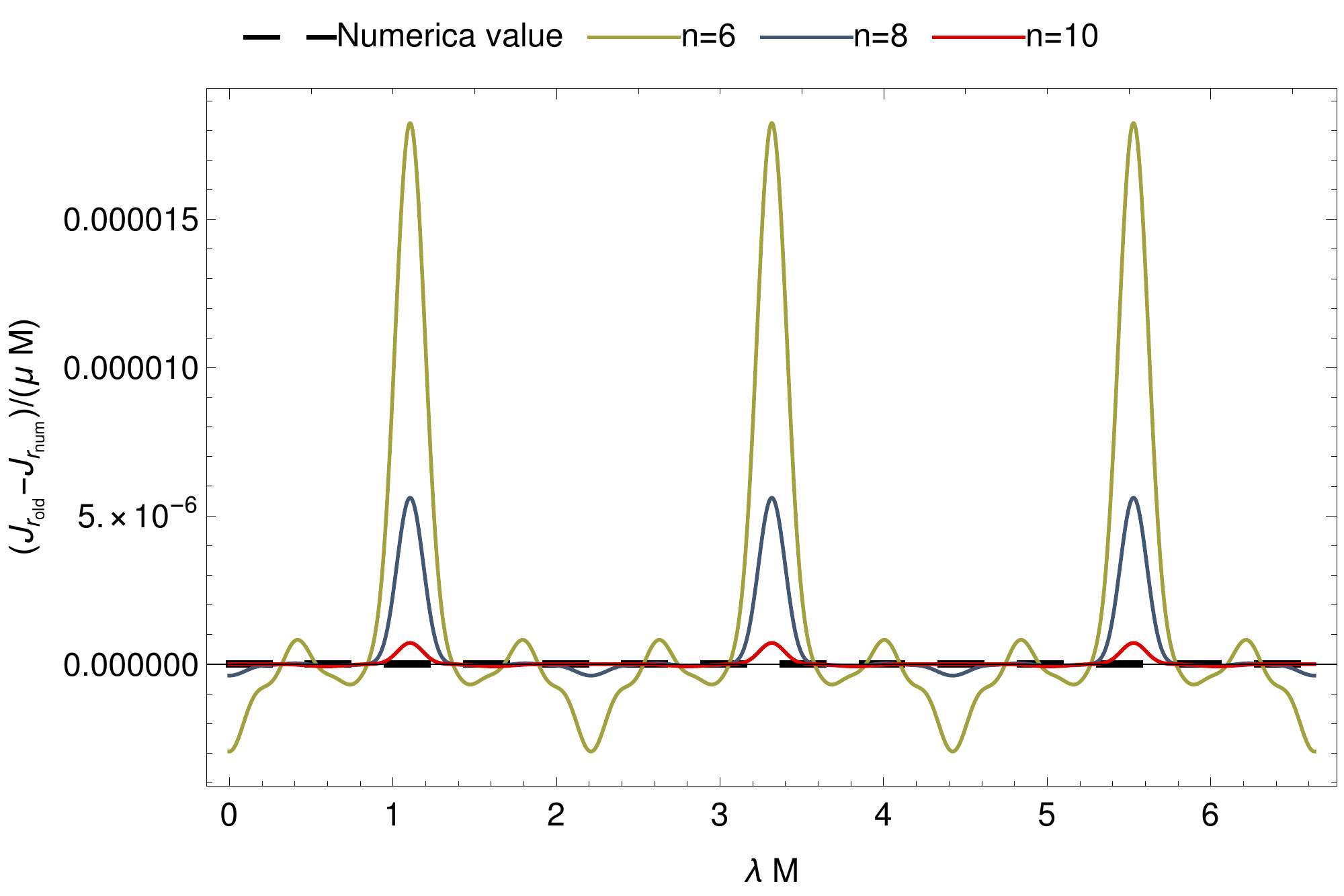}}
{  \subfloat[]{\includegraphics[width=0.48\textwidth]{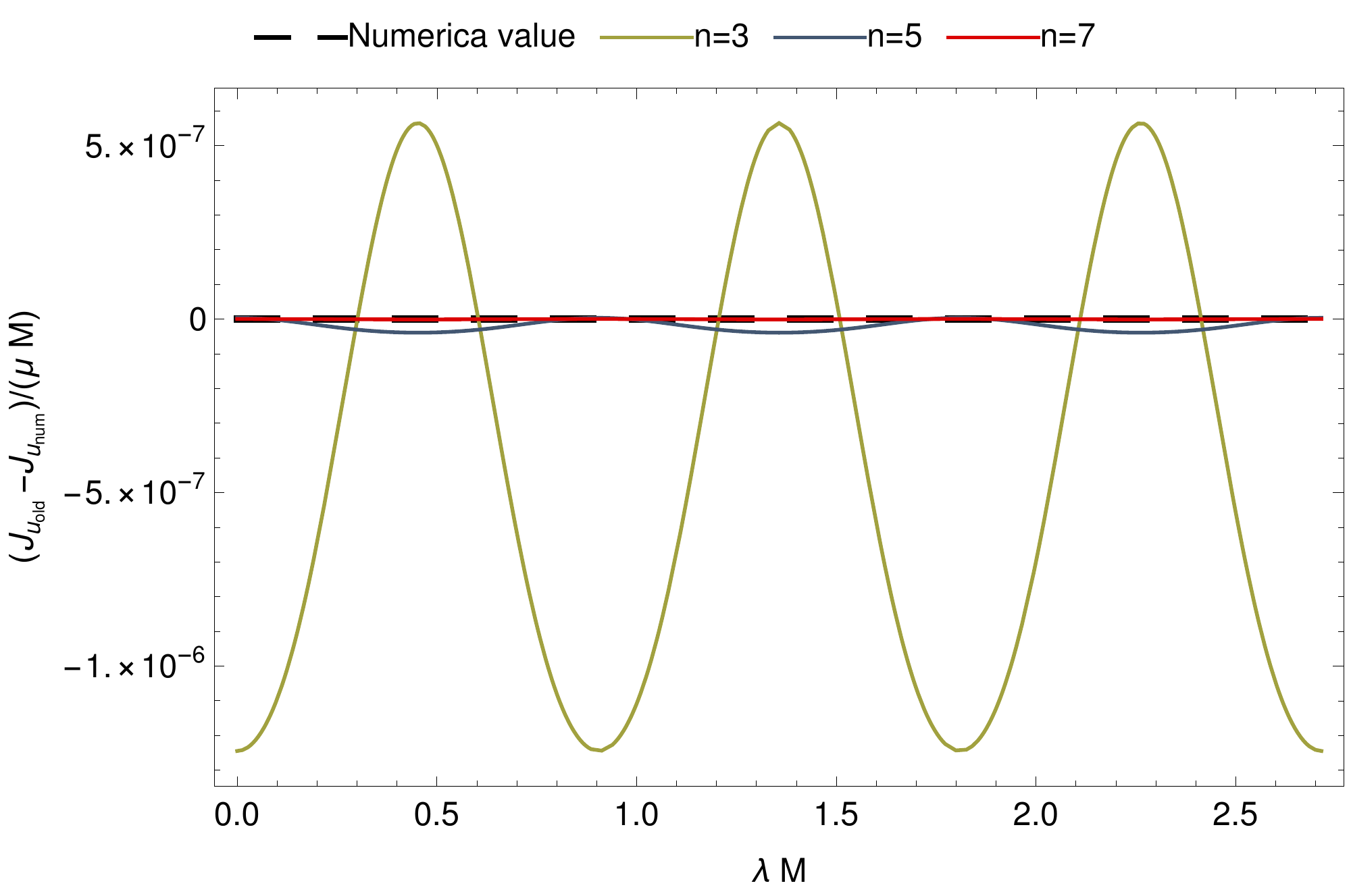}}}
}   
\end{center}
    \caption{These plots show by an increasing the number of transformations how the actions $J_{r_{\rm old}}$ and $J_{u_{\rm old}}$ are converging to their numerical values. For these figures we set  $a=0.99 M$, $e= 0.3$, semilatus rectum $p=10 M$, and initial inclination $\iota_0= \pi/8$. }\label{fig:new actions}
\end{figure*}

\begin{figure*}[ht]
\begin{center}
\captionsetup[subfloat]{position=top} 
 {\subfloat[]{\includegraphics[width=0.48\textwidth]{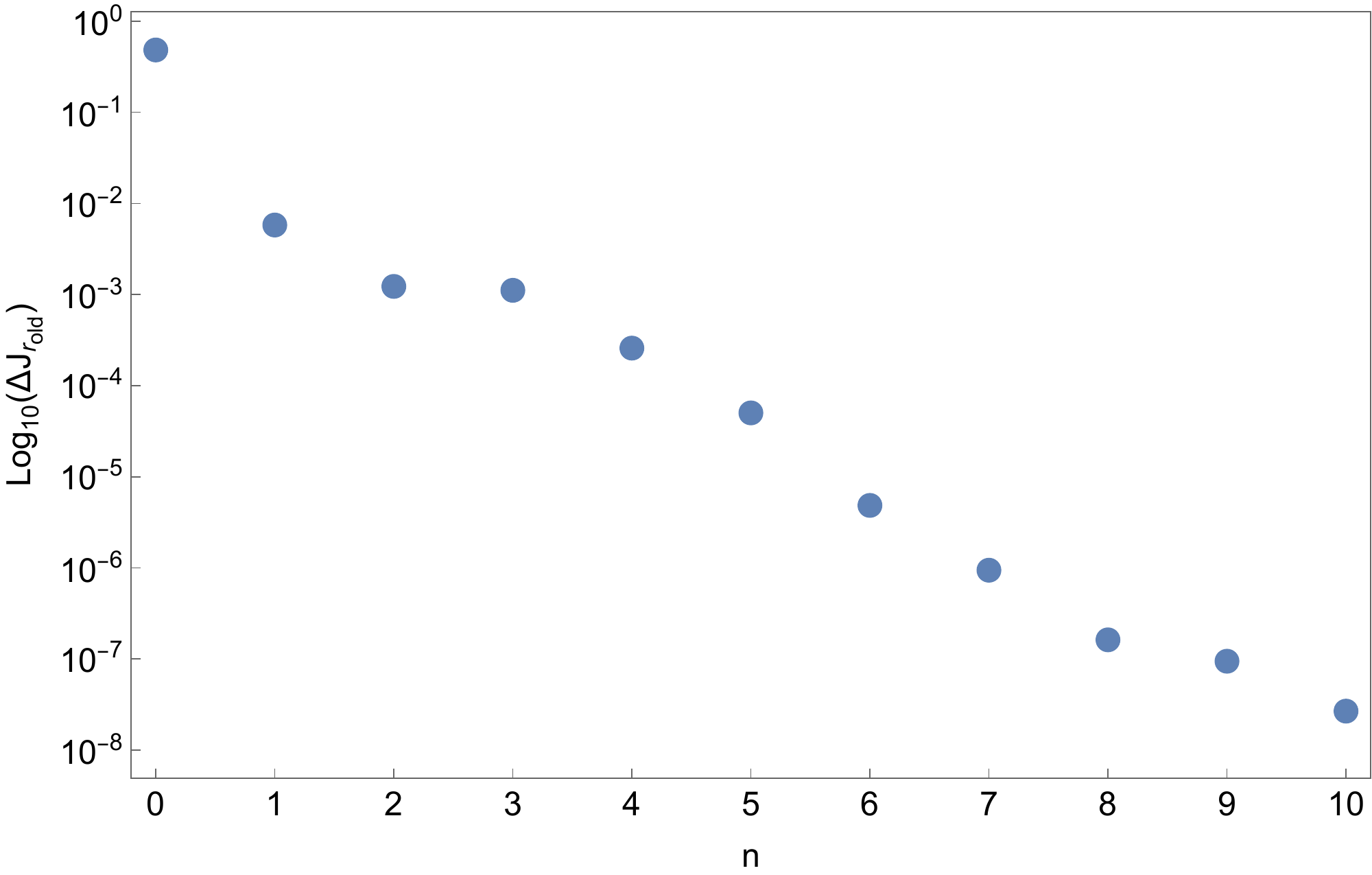}}
{  \subfloat[]{\includegraphics[width=0.48\textwidth]{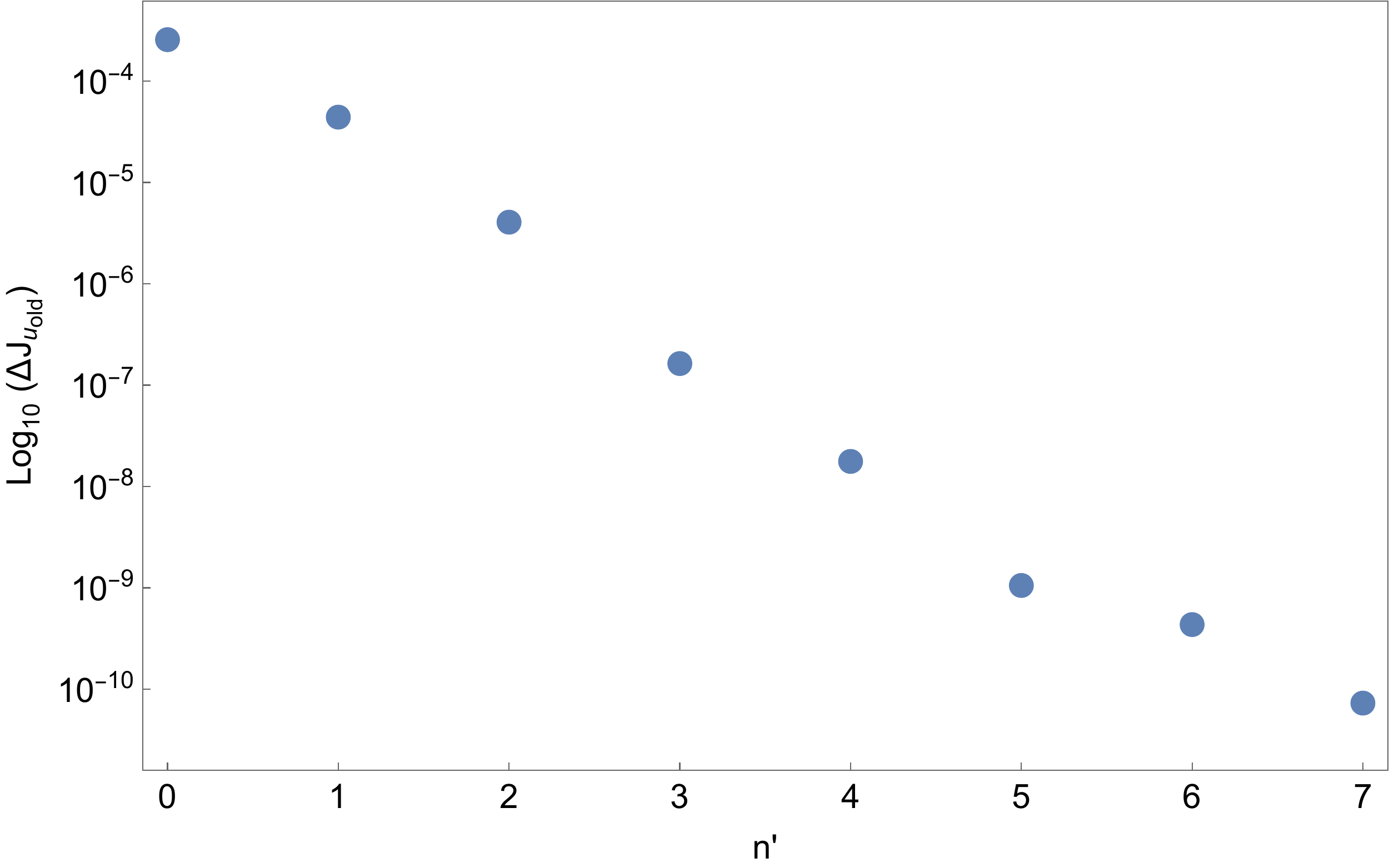}}}
}   
\end{center}
    \caption{These plots show $\Delta J_{r_{old}}$ and $\Delta J_{u_{old}}$, as defined in Eq.~\eqref{eq:defdelta}, at each order $n$ and $n^\prime$ respectively of a canonical transformation.  For these figures we set  $a=0.99 M$, $e= 0.3$, semilatus rectum $p=10 M$, and initial inclination $\iota_0= \pi/8$. }\label{fig:nthorder}
\end{figure*}

 We have $n$ generating functions for the radial transformation, i.e.  $\chi_{r_i}$ where  $i=\lbrace 1, n\rbrace$, and $n^\prime$ generating functions for the angular transformation, i.e. $\chi_{u_j}$ where $j= \lbrace 1,n^\prime \rbrace$. The crucial question is how can we determine the $n$ and $n^\prime$, namely how many times we should apply the canonical transformations? To find it out, we check if the new system $H_{AA}$ converges to the original system $H_{pn}$, namely, we see how accurate is the new system. Therefore, we check if the new actions $\lbrace J_t, J_r,J_u, J_{\nu} \rbrace$ are converging to their theoretical expected constant values. The $J_t$ and $J_{\nu}$ automatically converge, since the Hamiltonian does not explicitly depend on the respective angles.  However, for the $J_r$ and $J_u$ we have to write them in terms of the original variables by applying the inverse Lie transformations, in the same fashion as shown in Eq.~\eqref{eq:newtoold}, i.e.
\begin{align}
    J_{r_{\rm old}}& =\text{exp}(-\mathcal{L}_{\chi_{r_{1}}})\,\text{exp}(-\mathcal{L}_{\chi_{r_2}}) . . . \,\text{exp}(-\mathcal{L}_{\chi_{r_{n}}}) \,J_r,\label{eq:jrold}\\
    J_{u_{\rm old}}& =\text{exp}(-\mathcal{L}_{\chi_{u_{1}}})\,\text{exp}(-\mathcal{L}_{\chi_{u_2}}) . . . \,\text{exp}(-\mathcal{L}_{\chi_{u_{n^\prime}}}) \,J_u .\label{eq:juold}
\end{align}
In other words, by using the above transformations~\eqref{eq:jrold} and~\eqref{eq:juold}, we are able to express the new actions into the Boyer–Lindquist variables, i.e. $r(\lambda)$, $\theta(\lambda)$, $ L_z$ and etc.

As we mentioned earlier, the accuracy of the new system depends on the position of the reference orbit $r_c$ and how we deviate the system from the reference orbit, i.e. $\delta \,dr$. We observed that by choosing 
\begin{align}\label{eq:rcdelta}
    r_c&= \frac{p}{1-e^2}+ e  (1-10\,e) M, \nonumber \\
    \delta &= r_c \dfrac{a}{e\, M^2}
\end{align}
the new system obtains better accuracy. However, we should mention that there isn't any general prescription on how one should set $r_c$ and $\delta$. Note that our scheme diverges when the Kerr parameter tends to zero. In the Schwarzschild limit one should follow the approach suggested in \cite{Polcar22}.

In Figs.~\ref{fig:new actions}-\ref{fig:nthorder} we illustrate the aforementioned comparison for different $n$ and $n^\prime$. To illustrate the $J_{r_{\rm old}}$ and $J_{u_{\rm old}}$, we use the {\it KerrGeodesics} package of the  {\it Black Hole Perturbation Toolkit} repository~\cite{BHPT} since $J_{r_{\rm old}}$ and $J_{u_{\rm old}}$ are functions of  $r(\lambda)$ and $\theta(\lambda)$ respectively. The numerical values for $\overline{J}_r$ and $\overline{J}_u$ are computed from~\eqref{eq:jrpl}-~\eqref{eq:jupl} where we numerically integrate them for given values of $E$, $L_z,$ and $Q$ from the parameters $\lbrace a, p, e, \iota_0\rbrace$~\cite{Schmidt02,Drasco06}. Figs.~\ref{fig:new actions}-\ref{fig:nthorder} show that by increasing the number of canonical transformations, the actions converge to their numerical values, indicating that the approximate system is converging to the original one. More precisely, applying more and more canonical transformations reduces the oscillations of the actions around their numerical values (see Fig.~\ref{fig:new actions}), while for the number of transformations shown in Fig.~\ref{fig:nthorder} the convergence appears to be exponential. Although the actions are converging, they have still a small oscillation around the value they converge to.

As mentioned in Sec.~\ref{sec:intro}, modeling an EMRI system requires accuracy below $10^{-5}$ during one cycle. Thus, in our approximation system, we require to have an error not greater than $10^{-5}$. Therefore, we set $10$ canonical transformations for the radial part and $7$ canonical transformations for the angular part. For these numbers of canonical transformations, the oscillations of the actions compared to the numerical values are small enough. The accuracy of the radial part by setting $n=10$ is in the range $\sim 10^{-11}- 10^{-9}$ for small eccentricity, i.e. $\sim 0.1$, and by increasing the eccentricity the accuracy is decreasing to $\sim 10^{-5}$ for eccentricity $\sim 0.5$; we observed that by increasing the numbers of canonical transformations for the radial part, the accuracy does not improve significantly. However, for the angular part for $n^\prime = 7$ we get the accuracy $\sim 10^{-14}- 10^{-9}$ for small eccentricities $e \sim 0.1$, while when the eccentricity is  approaching to $0.5$ the accuracy drops to $\sim 10^{-11}- 10^{-8}$.  For a fixed eccentricity, the accuracy decreases as the Kerr parameter increases, while increasing the initial inclination does not have much influence on the accuracy of the system. Tables~\ref{tab:rp10i8} and~\ref{tab:rp30i3} show the accuracy of the resulting system for the radial part and tables~\ref{tab:up10i8} and~\ref{tab:up30i3} show the accuracy achieved for the angular part for two different initial conditions. Table~\ref{tab:changeincl} shows a case of how the angular part depends on the inclination. In these tables, the $\Delta$ and $\delta$ correspond to the relative errors and are defined as
\begin{align}\label{eq:defdelta}
     \Delta X= \lvert 1- \frac{X_{\rm max}}{X_{\rm num}}\rvert,  \qquad \delta X= \lvert 1- \frac{X_{\rm min}}{X_{\rm max}} \rvert,
\end{align}
where the $X_{\rm max}$ and $X_{\rm min}$ are the maximum and the minimum values of the oscillation of each variable $X$ and $X_{\rm num}$ denote its numerical value. In these tables, the actions are computed from Eqs.~\eqref{eq:jrold} and~\eqref{eq:juold} and $\Upsilon_i= \partial H_{AA}/ \partial J_i$; the numerical values of the frequencies $\Upsilon_{i_{\rm num}}$ are obtained from the {\it KerrGeodesics} package of the  {\it Black Hole Perturbation Toolkit} repository. 

\begin{table}[]
\caption{ Here we provide the order of the relative errors for quantities related with the radial motion. $\Delta J_{r_{\rm old}}$ represents the relative error of the maximum value of the $J_{r_{\rm old}}$ with respect to its numerical values; $\delta J_{r_{\rm old}}$ shows the relative error between the maximum and minimum values of the $J_{\rm old}$; and the $\Delta \Upsilon_r$ shows the relative error of the maximum $\Upsilon_r$ and its numerical value, i.e. $\Upsilon_{r_{\rm num}}$. The orbit has semi-latus rectum $p=10 M$ and inclination $\iota_0=\frac{\pi}{8}$, and to approximate the system $10$ canonical transformations have been applied. See Sec.~\ref{sec:NewHam} for more details.  }
\label{tab:rp10i8}
\begin{tabular}{|l|c|c|c|c|c|}
\hline
\multicolumn{1}{|c|}{$a$}     & $e$&$\mathcal{O}(\Delta J_{r_{\rm old}})$&$\mathcal{O}(\delta J_{r_{\rm old}})$ &$\mathcal{O}(\Delta \Upsilon_r)$\\ \hline
                              & $0.1$& $10^{-9 }$            &$10^{-9 }$                & $10^{-11}$              \\\cline{2-5}
\multicolumn{1}{|c|}{$0.1$}   & $0.2$& $10^{-8 }$            &$10^{-8 }$                & $10^{-9 }$              \\\cline{2-5}
                              & $0.3$& $10^{-6 }$            &$10^{-6 }$                & $10^{-7 }$              \\\cline{2-5}
                              & $0.4$& $10^{-5 }$            &$10^{-5 }$                & $10^{-5 }$              \\\cline{2-5}
                              & $0.5$& $10^{-4 }$            &$10^{-4 }$                & $10^{-4 }$              \\\hline
                              & $0.1$& $10^{-10}$            &$10^{-10}$                & $10^{-12}$              \\\cline{2-5}
\multicolumn{1}{|c|}{$0.3$}   & $0.2$& $10^{-8 }$            &$10^{-8 }$                & $10^{-9 }$              \\\cline{2-5}
                              & $0.3$& $10^{-6 }$            &$10^{-6 }$                & $10^{-7 }$              \\\cline{2-5} 
                              & $0.4$& $10^{-5 }$            &$10^{-5 }$                & $10^{-6 }$              \\\cline{2-5}
                              & $0.5$& $10^{-4 }$            &$10^{-4 }$                & $10^{-5 }$              \\ \hline
                              & $0.1$& $10^{-9 }$            &$10^{-9 }$                & $10^{-11}$              \\\cline{2-5}
\multicolumn{1}{|c|}{$0.5$}   & $0.2$& $10^{-9 }$            &$10^{- 7}$                & $10^{-10}$              \\\cline{2-5}
                              & $0.3$& $10^{- 7}$            &$10^{-7 }$                & $10^{-8 }$              \\\cline{2-5}
                              & $0.4$& $10^{- 5}$            &$10^{-5 }$                & $10^{-6 }$              \\\cline{2-5}
                              & $0.5$& $10^{-4 }$            &$10^{-4 }$                & $10^{-5 }$              \\ \hline
                              & $0.1$& $10^{-7 }$            &$10^{-7 }$                & $10^{-10}$              \\\cline{2-5}
\multicolumn{1}{|c|}{$0.7$}   & $0.2$& $10^{-7 }$            &$10^{-7 }$                & $10^{-8 }$              \\\cline{2-5}
                              & $0.3$& $10^{-7 }$            &$10^{-6 }$                & $10^{-8 }$              \\\cline{2-5}
                              & $0.4$& $10^{-6 }$            &$10^{-5 }$                & $10^{-7 }$              \\\cline{2-5}
                              & $0.5$& $10^{-5 }$            &$10^{-5 }$                & $10^{-5 }$              \\ \hline
                              & $0.1$& $10^{-7 }$            &$10^{-7 }$                & $10^{-10}$              \\\cline{2-5}
\multicolumn{1}{|c|}{$0.99$}  & $0.2$& $10^{-6 }$            &$10^{-6 }$                & $10^{-8 }$              \\\cline{2-5}
                              & $0.2$& $10^{-7 }$            &$10^{-7 }$                & $10^{-8 }$              \\\cline{2-5}
                              & $0.4$& $10^{-6 }$            &$10^{-6 }$                & $10^{-5 }$              \\\cline{2-5}
                              & $0.5$& $10^{-5 }$            &$10^{-5 }$                & $10^{-5 }$              \\ \hline
\end{tabular}
\end{table}

\begin{table}[]
\caption{ As in Table~\ref{tab:rp10i8}, but for an orbit with semilatus rectum $p=30 M$ and inclination $\iota_0=\frac{\pi}{3}$.}
\label{tab:rp30i3}
\begin{tabular}{|l|c|c|c|c|c|}
\hline
\multicolumn{1}{|c|}{$a$}     & $e$  & $\mathcal{O}(\Delta J_{r_{\rm old}})$&$\mathcal{O}(\delta J_{r_{\rm old}})$&$\mathcal{O}(\delta \Upsilon_r)$\\ \hline
                              & $0.1$&$10^{-11}$                & $10^{-11}$              &$10^{-10}$\\\cline{2-5}
\multicolumn{1}{|c|}{$0.1$}   & $0.2$&$10^{-8 }$                & $10^{-8 }$              &$10^{-7 }$\\\cline{2-5}
                              & $0.3$&$10^{-6 }$                & $10^{-6 }$              &$10^{-6 }$\\\cline{2-5}
                              & $0.4$&$10^{-5 }$                & $10^{-5 }$              &$10^{-5 }$\\\cline{2-5}
                              & $0.5$&$10^{-3 }$                & $10^{-3 }$              &$10^{-4 }$\\\hline
                              & $0.1$&$10^{-7 }$                & $10^{-7 }$              &$10^{-10 }$\\\cline{2-5}
\multicolumn{1}{|c|}{$0.3$}   & $0.2$&$10^{-8 }$                & $10^{-8 }$              &$10^{-8 }$\\\cline{2-5}
                              & $0.3$&$10^{-6 }$                & $10^{-6 }$              &$10^{-6 }$\\\cline{2-5}
                              & $0.4$&$10^{-5 }$                & $10^{-5 }$              &$10^{-6 }$\\\cline{2-5}
                              & $0.5$&$10^{-3 }$                & $10^{-3 }$              &$10^{-4 }$\\ \hline
                              & $0.1$&$10^{-7 }$                & $10^{-7 }$              &$10^{-10 }$\\\cline{2-5}
\multicolumn{1}{|c|}{$0.5$}   & $0.2$&$10^{- 7}$                & $10^{- 8}$              &$10^{-8 }$\\\cline{2-5}
                              & $0.3$&$10^{-6 }$                & $10^{-7 }$              &$10^{-6 }$\\\cline{2-5}
                              & $0.4$&$10^{-5 }$                & $10^{-5 }$              &$10^{-5 }$\\\cline{2-5}
                              & $0.5$&$10^{-3 }$                & $10^{-3 }$              &$10^{-4 }$\\ \hline
                              & $0.1$&$10^{-7 }$                & $10^{-7 }$              &$10^{-9 }$\\\cline{2-5}
\multicolumn{1}{|c|}{$0.7$}   & $0.2$&$10^{-6 }$                & $10^{-6 }$              &$10^{-7 }$\\\cline{2-5}
                              & $0.3$&$10^{-7 }$                & $10^{-7 }$              &$10^{-7 }$\\\cline{2-5}
                              & $0.4$&$10^{-5 }$                & $10^{-5 }$              &$10^{-6 }$\\\cline{2-5}
                              & $0.5$&$10^{-3 }$                & $10^{-3 }$              &$10^{-4 }$\\ \hline
                              & $0.1$&$10^{-5 }$                & $10^{-5 }$              &$10^{-8 }$\\\cline{2-5}
\multicolumn{1}{|c|}{$0.99$}  & $0.2$&$10^{-6 }$                & $10^{-6 }$              &$10^{-7 }$\\\cline{2-5}
                              & $0.3$&$10^{-5 }$                & $10^{-5 }$              &$10^{-6 }$\\\cline{2-5}
                              & $0.4$&$10^{-4 }$                & $10^{-4 }$              &$10^{-5 }$\\\cline{2-5}
                              & $0.5$&$10^{-3 }$                & $10^{-3 }$              &$10^{-4 }$\\ \hline
\end{tabular}
\end{table}

\begin{table}[]
\caption{  Here we provide the order of the relative errors for quantities related to the angular motion. $\Delta J_{u_{\rm old}}$ represents the relative error of the maximum value of the $J_{u_{\rm old}}$ with respect to its numerical values; $\delta J_{u_{\rm old}}$ shows the relative error between the maximum and minimum values of the $J_{u_{\rm old}}$; and the $\Delta \Upsilon_u$ shows the relative error of the maximum $\Upsilon_u$ and its numerical value, i.e. $\Upsilon_{u_{\rm num}}$. The orbit has semilatus rectum
 $p=10 M$ and inclination $\iota_0=\frac{\pi}{8}$ to approximate the system $7$ canonical transformations have been applied. See Sec.~\ref{sec:NewHam} for more details.  }
\label{tab:up10i8}
\begin{tabular}{|l|c|c|c|c|c|}
\hline
\multicolumn{1}{|c|}{$a$}     & $e$  &$\mathcal{O}(\Delta \Upsilon_u)$& $\mathcal{O}(\Delta Q)$& $\mathcal{O}(\Delta J_{u_{\rm old}})$&$\mathcal{O}(\delta J_{u_{\rm old}})$\\ \hline
                              & $0.1$& $10^{-11}$            & $10^{-10}$             &$10^{-9 }$                & $10^{-10}$              \\\cline{2-6}
\multicolumn{1}{|c|}{$0.1$}   & $0.2$& $10^{-11}$            & $10^{-10}$             &$10^{-9 }$                & $10^{-10}$              \\\cline{2-6}
                              & $0.3$& $10^{-11}$            & $10^{-10}$             &$10^{-10}$                & $10^{-12}$              \\\cline{2-6}
                              & $0.4$& $10^{-11}$            & $10^{-10}$             &$10^{-10}$                & $10^{-11}$              \\\cline{2-6}
                              & $0.5$& $10^{-11}$            & $10^{-10}$             &$10^{-10}$                & $10^{-10}$              \\\hline
                              & $0.1$& $10^{-12}$            & $10^{-11}$             &$10^{-11}$                & $10^{-10}$              \\\cline{2-6}
\multicolumn{1}{|c|}{$0.3$}   & $0.2$& $10^{-12}$            & $10^{-11}$             &$10^{-11}$                & $10^{-11}$              \\\cline{2-6}
                              & $0.3$& $10^{-12}$            & $10^{-11}$             &$10^{-11}$                & $10^{-11}$              \\\cline{2-6}
                              & $0.4$& $10^{-11}$            & $10^{-11}$             &$10^{-11}$                & $10^{-11}$              \\\cline{2-6}
                              & $0.5$& $10^{-10}$            & $10^{-11}$             &$10^{-11}$                & $10^{-11}$              \\\hline
                              & $0.1$& $10^{-11}$            & $10^{-10}$             &$10^{-10}$                & $10^{-10}$              \\\cline{2-6}
\multicolumn{1}{|c|}{$0.5$}   & $0.2$& $10^{-11}$            & $10^{-10}$             &$10^{-10}$                & $10^{-10}$              \\\cline{2-6}
                              & $0.3$& $10^{-11}$            & $10^{-10}$             &$10^{-10}$                & $10^{-10}$              \\\cline{2-6}
                              & $0.4$& $10^{-10}$            & $10^{-10}$             &$10^{-10}$                & $10^{-10}$              \\\cline{2-6}
                              & $0.5$& $10^{-9 }$            & $10^{-10}$             &$10^{-10}$                & $10^{-10}$              \\\hline
                              & $0.1$& $10^{-10}$            & $10^{-10}$             &$10^{-10}$                & $10^{-10}$              \\\cline{2-6}
\multicolumn{1}{|c|}{$0.7$}   & $0.2$& $10^{-10}$            & $10^{-10}$             &$10^{-10}$                & $10^{-10}$              \\\cline{2-6}
                              & $0.3$& $10^{-10}$            & $10^{-10}$             &$10^{-10}$                & $10^{-10}$              \\\cline{2-6}
                              & $0.4$& $10^{-9 }$            & $10^{-10}$             &$10^{-10}$                & $10^{-10}$              \\\cline{2-6}
                              & $0.5$& $10^{-9 }$            & $10^{-10}$             &$10^{-10}$                & $10^{-10}$              \\\hline
                              & $0.1$& $10^{-9 }$            & $10^{-10}$             &$10^{-9 }$                & $10^{-9 }$              \\\cline{2-6}
\multicolumn{1}{|c|}{$0.99$}  & $0.2$& $10^{-9 }$            & $10^{-10}$             &$10^{-9 }$                & $10^{-9 }$              \\\cline{2-6}
                              & $0.3$& $10^{-10}$            & $10^{-9 }$             &$10^{-9 }$                & $10^{-9 }$              \\\cline{2-6}
                              & $0.4$& $10^{-9 }$            & $10^{-9 }$             &$10^{-9 }$                & $10^{-9 }$              \\\cline{2-6}
                              & $0.5$& $10^{-8 }$            & $10^{-9 }$             &$10^{-9 }$                & $10^{-9 }$              \\\hline
\end{tabular}
\end{table}

\begin{table}[]
\caption{ As in Table~\ref{tab:up10i8}, but for an orbit with semilatus rectum $p=30 M$ and inclination $\iota_0=\frac{\pi}{3}$.}
\label{tab:up30i3}
\begin{tabular}{|l|c|c|c|c|c|}
\hline
\multicolumn{1}{|c|}{$a$}     & $e$  &$\mathcal{O}(\Delta \Upsilon_u)$& $\mathcal{O}(\Delta Q)$& $\mathcal{O}(\Delta J_{u_{\rm old}})$&$\mathcal{O}(\delta J_{u_{\rm old}})$\\ \hline
                              & $0.1$& $10^{-12}$            & $10^{-12}$             &$10^{-10}$                & $10^{-10}$              \\\cline{2-6}
\multicolumn{1}{|c|}{$0.1$}   & $0.2$& $10^{-12}$            & $10^{-12}$             &$10^{-10}$                & $10^{-10}$              \\\cline{2-6}
                              & $0.3$& $10^{-12}$            & $10^{-12}$             &$10^{-11}$                & $10^{-11}$              \\\cline{2-6}
                              & $0.4$& $10^{-12}$            & $10^{-12}$             &$10^{-11}$                & $10^{-11}$              \\\cline{2-6}
                              & $0.5$& $10^{-11}$            & $10^{-12}$             &$10^{-11}$                & $10^{-11}$              \\\hline 
                              & $0.1$& $10^{-14}$            & $10^{-14}$             &$10^{-12}$                & $10^{-13}$              \\\cline{2-6}
\multicolumn{1}{|c|}{$0.3$}   & $0.2$& $10^{-11}$            & $10^{-12}$             &$10^{-11}$                & $10^{-11}$              \\\cline{2-6}
                              & $0.3$& $10^{-12}$            & $10^{-11}$             &$10^{-11}$                & $10^{-11}$              \\\cline{2-6}
                              & $0.4$& $10^{-11}$            & $10^{-11}$             &$10^{-11}$                & $10^{-11}$              \\\cline{2-6}
                              & $0.5$& $10^{-10}$            & $10^{-11}$             &$10^{-10}$                & $10^{-10}$              \\\hline 
                              & $0.1$& $10^{-13}$            & $10^{-13}$             &$10^{-12}$                & $10^{-12}$              \\\cline{2-6}
\multicolumn{1}{|c|}{$0.5$}   & $0.2$& $10^{-13}$            & $10^{-13}$             &$10^{-12}$                & $10^{-12}$              \\\cline{2-6}
                              & $0.3$& $10^{-13}$            & $10^{-13}$             &$10^{-12}$                & $10^{-12}$              \\\cline{2-6}
                              & $0.4$& $10^{-11}$            & $10^{-13}$             &$10^{-13}$                & $10^{-11}$              \\\cline{2-6}
                              & $0.5$& $10^{-10}$            & $10^{-11}$             &$10^{-11}$                & $10^{-10}$              \\\hline 
                              & $0.1$& $10^{-12}$            & $10^{-12}$             &$10^{-11}$                & $10^{-11}$              \\\cline{2-6}
\multicolumn{1}{|c|}{$0.7$}   & $0.2$& $10^{-12}$            & $10^{-12}$             &$10^{-11}$                & $10^{-11}$              \\\cline{2-6}
                              & $0.3$& $10^{-12}$            & $10^{-12}$             &$10^{-12}$                & $10^{-11}$              \\\cline{2-6}
                              & $0.4$& $10^{-11}$            & $10^{-11}$             &$10^{-11}$                & $10^{-11}$              \\\cline{2-6}
                              & $0.5$& $10^{-10}$            & $10^{-11}$             &$10^{-10}$                & $10^{-10}$              \\\hline 
                              & $0.1$& $10^{-11}$            & $10^{-11}$             &$10^{-10}$                & $10^{-10}$              \\\cline{2-6}
\multicolumn{1}{|c|}{$0.99$}  & $0.2$& $10^{-11}$            & $10^{-11}$             &$10^{-10}$                & $10^{-10}$              \\\cline{2-6}
                              & $0.3$& $10^{-11}$            & $10^{-11}$             &$10^{-11}$                & $10^{-11}$              \\\cline{2-6}
                              & $0.4$& $10^{-11}$            & $10^{-12}$             &$10^{-11}$                & $10^{-11}$              \\\cline{2-6}
                              & $0.5$& $10^{-10}$            & $10^{-10}$             &$10^{-10}$                & $10^{-10}$              \\\hline 
\end{tabular}
\end{table}

\begin{table}[]
\caption{As in Table~\ref{tab:up10i8}, but for $\lbrace a, p, e \rbrace= \lbrace 0.99, 10, 0.5 \rbrace $ and different inclinations.}
\label{tab:changeincl}
\begin{tabular}{|l|c|c|c|c|}
\hline
 $\iota$  &$\mathcal{O}(\Delta \Upsilon_u)$& $\mathcal{O}(\Delta Q)$& $\mathcal{O}(\Delta J_{u_{\rm old}})$&$\mathcal{O}(\delta J_{u_{\rm old}})$\\ \hline
 $\pi/8$      & $10^{-8 }$            & $10^{-9 }$             &$10^{-9 }$                & $10^{-9 }$    \\ \hline         
 $\pi/7$      & $10^{-8 }$            & $10^{-9 }$             &$10^{-9 }$                & $10^{-9 }$   \\ \hline
 $\pi/6$      & $10^{-8 }$            & $10^{-9 }$             &$10^{-9 }$                & $10^{-9 }$   \\ \hline
 $\pi/5$      & $10^{-8 }$            & $10^{-9 }$             &$10^{-9 }$                & $10^{-9 }$   \\ \hline
 $\pi/4$      & $10^{-8 }$            & $10^{-9 }$             &$10^{-9 }$                & $10^{-9 }$   \\ \hline
 $\pi/3$      & $10^{-9 }$            & $10^{-9 }$             &$10^{-10}$                & $10^{-9 }$   \\ \hline
                     
\end{tabular}
\end{table}

\subsection{Perturbed-Kerr trajectory}\label{sec:pertbtraj}

Up to here, we have just expressed up to a certain accuracy the Hamiltonian in the AA variables and the respective generating functions have been derived. In what follows we use the acquired system to generate the trajectories in AA variables. 

In our approximative formalism, the evolution equations are given by
\begin{align}
    \Upsilon_r &=\dot{ \psi_r}= \frac{\partial H_{AA}}{\partial J_r},\label{eq:rfrec}\\
    \Upsilon_u &= \dot{\psi_u}= \frac{\partial H_{AA}}{\partial J_u},\label{eq:ufrec}\\
    \Upsilon_\nu &= \dot{\psi_\nu}= \frac{\partial H_{AA}}{\partial J_{\nu}},\label{eq:nufrec}\\
    \Upsilon &= \dot{\psi_t}= \frac{\partial H_{AA}}{\partial J_t},\label{eq:tfrec}
\end{align}
where $ \Upsilon_i$ denotes the approximative Mino-time frequency and over-dot denotes the derivative with respect to the Mino time. The transformation~\eqref{eq:cantrans} implies that $\Upsilon_u=\Upsilon_\theta$ for the Schwarzschild case. For the Kerr case, it's not so obvious that this relation should hold, we have confirmed, however, numerically that it does hold. 

From the above equations of motion, we derive the trajectories in the AA phase space $\lbrace J_r, J_u, J_{\nu}, J_t, \psi_r, \psi_u, \psi_\nu, \psi_t \rbrace$. 
We apply the Lie series~\eqref{eq:oldtonew} to transform the above approximative trajectories in terms of the Boyer-Lindquist coordinates. In order to achieve this, we start from the point at which no Lie series has been applied yet, i.e. from the Eqs.~\eqref{eq:cantrans},~\eqref{eq:bkkp},~\eqref{eq:harmonictrans}, in which we replaced $u$ with ${\psi_u}^{0}$ and $\nu$ with ${\psi_\nu}^{0}$. Namely, in that step, the transformation is given by
\begin{align}
    r_{\text{old}}&= r_c+ \sqrt{\frac{2\,J_{r}^{0}}{m_c\, \Omega_{r0}}} \sin \psi_{r}^{0},\label{eq:rold}\\
    \theta_{\text{old}}&= \arccos \left(\sqrt{1-\frac{(J_{\nu}^0+L_{zc})^2}{(J_{u}^{0}+p_{uc})^2}} \sin \psi_u^0\right),\label{eq:thold}\\
     \phi_{\text{old}}&=\psi_\nu^0+\psi_u^0\nonumber\\
     & + \arctan\left(\frac{\left(\frac{J_{\nu}^0+L_{zc}}{J_{u}^0+p_{uc}}-1\right) \cos \psi_u^0 \sin \psi_u^0}{1+ \left(\frac{J_{\nu}^0+L_{zc}}{J_{u}^0+p_{uc}}-1\right) \sin^2 \psi_u^0} \right),\label{eq:phiold}\\
     t_{\rm old}&= \psi_t^0\label{eq:told}.
\end{align}
Then the approximative trajectories in Boyer-Lindquist coordinates are determined from 
\begin{align} 
r_{\text{new}}=&[\text{exp}(\mathcal{L}_{\chi_{r_{10}}})\,\text{exp}(\mathcal{L}_{\chi_{r_{9}}})...\text{exp}(\mathcal{L}_{\chi_{r_1}})] \,r_{\text{old}},\label{eq:rnew}\\
     \theta_{\text{new}}=&[\text{exp}(\mathcal{L}_{\chi_{u_{7}}})\,\text{exp}(\mathcal{L}_{\chi_{u_{6}}})...\text{exp}(\mathcal{L}_{\chi_{u_1}})] \,\theta_{\text{old}}\label{eq:thnew},\\
\phi_{\text{new}}=& [\text{exp}(\mathcal{L}_{\chi_{r_{10}}})\,\text{exp}(\mathcal{L}_{\chi_{r_{9}}})...\text{exp}(\mathcal{L}_{\chi_{r_1}})]\nonumber \\ \times&[\text{exp}(\mathcal{L}_{\chi_{u_{7}}})\,\text{exp}(\mathcal{L}_{\chi_{u_{6}}})...\text{exp}(\mathcal{L}_{\chi_{u_1}}) ]\,\phi_{\text{old}},\label{eq:phinew}\\
t_{\text{new}}=& [\text{exp}(\mathcal{L}_{\chi_{r_{10}}})\,\text{exp}(\mathcal{L}_{\chi_{r_{9}}})...\text{exp}(\mathcal{L}_{\chi_{r_1}})]\nonumber\\ \times&[\text{exp}(\mathcal{L}_{\chi_{u_{7}}})\,\text{exp}(\mathcal{L}_{\chi_{u_{6}}})...\text{exp}(\mathcal{L}_{\chi_{u_1}})]\,t_{\text{old}}\label{eq:tnew}.
\end{align}
By applying the above Lie series transformations we essentially express $r_{\text{new}},~\theta_{\text{new}},~\phi_{\text{new}},~t_{\text{new}}$ as functions of $\lbrace J_r, J_u, J_{\nu}, J_t, \psi_r, \psi_u, \psi_\nu, \psi_t \rbrace$. Note that, since the $\chi_{u_i}$ are independent of $J_r^0$ and $\psi_r^0$, Eq.~\eqref{eq:rnew} is independent of ${\text{exp}}(\mathcal{L}_{\chi_{u_i}})$. In the same fashion Eq.~\eqref{eq:thnew} is independent of ${\text{exp}}(\mathcal{L}_{\chi_{r_i}})$. On the other hand, both $\chi_{u_i}$ and $\chi_{r_i}$ depend on $J_t^0$ and $J_{\nu}^0$, that is why we apply both Lie series transformations in the case of  Eqs.~\eqref{eq:phinew},~\eqref{eq:tnew}. In the latter case, the resulting equations for $\phi_{\rm new}$ and $t_{\rm new}$ have the following form   
\begin{align}
    \phi_{\rm new}&=\psi_\nu+\psi_u+ \Delta \phi_r [\chi_{r_i}]+  \Delta \phi_\theta [\chi_{u_i}],\label{eq:phinewf}\\
   t_{\rm new}&= \psi_t+ \Delta t_r [\chi_{r_i}]+  \Delta t_\theta [\chi_{u_i}],\label{eq:tnewf}
\end{align}
where $\Delta \phi_r [\chi_{r_i}]$ and $\Delta t_r [\chi_{r_i}]$  correspond to the parts of $\phi_{\rm new}$ and $t_{\rm new}$ derived from applying the radial generating functions $\chi_{r_i}$; and the  $\Delta\phi_\theta [\chi_{u_i}]$ and $\Delta t_\theta [\chi_{u_i}]$ are determined from applying the angular generating functions $\chi_{u_i}$. Equations~\eqref{eq:phinewf} and~\eqref{eq:tnewf} are similar to the expressions given in Ref.~\cite{Hughes21}
\begin{align}
       \phi(\lambda)&=\phi_0+ \overline{\Upsilon}_\phi \lambda + \Delta \phi_r [r(\lambda)]+  \Delta \phi_\theta [\theta (\lambda)],\label{eq:phidras}\\
   t(\lambda)&= t_0 + \overline{\Upsilon} \lambda
   + \Delta t_r [r(\lambda)]+  \Delta t_\theta [\theta (\lambda)],\label{eq:tnewdras} 
\end{align}
where the over-line refers to the exact (not approximated) value of the frequencies. By comparing the Eq.~\eqref{eq:phidras} with the Eq.~\eqref{eq:phinewf} we conclude that $\psi_\phi=\psi_\nu+\psi_u$ and this indicates that the orbital plane precession frequency is $\Upsilon_\nu=\Upsilon_\phi -\Upsilon_u$.

 \section{Gravitational wave fluxes and generic adiabatic inspiral} \label{sec:inspiral}

\begin{table*}[]
 \caption{ Table of constants of motion fluxes through the horizon and to infinity for $a=0.9M$, $p=6M$. }
\centering
 \label{tab:fluxes}
\begin{tabular}{|l|c|c|c|c|c|c|c|}
\hline
\multicolumn{1}{|c|}{$e$}     & $\iota_0$&$(M/\mu)^2\Big\langle\frac{\mathrm{d}E^{\infty}}{\mathrm{d}t}\Big\rangle$&$(M/\mu)^2\Big\langle\frac{\mathrm{d}E^{H}}{\mathrm{d}t}\Big\rangle$ &$M/\mu^2\Big\langle\frac{\mathrm{d}L_z^{\infty}}{\mathrm{d}t}\Big\rangle$ &$M/\mu^2\Big\langle\frac{\mathrm{d}L_z^{H}}{\mathrm{d}t}\Big\rangle$  &$1/(M\mu^2)\Big\langle\frac{\mathrm{d}Q^{\infty}}{\mathrm{d}t}\Big\rangle$ &$1/(M\mu^2)\Big\langle\frac{\mathrm{d}Q^{H}}{\mathrm{d}t}\Big\rangle$\\ \hline
                              & $20 {}^{\circ}$& $5.87342\times 10^{-4}$   &$-4.25247\times 10^{-6}$  & $8.53698\times 10^{-3}$  & $-6.71479\times 10^{-5}$ & $5.24007\times 10^{-3}$ & $-3.30062\times 10^{-5}$  \\\cline{2-8}
\multicolumn{1}{|c|}{$0.1$}   & $40 {}^{\circ}$& $6.18311\times 10^{-4}$   &$-3.94869\times 10^{-6}$  & $7.63084\times 10^{-3}$  & $-7.72832\times 10^{-5}$ & $2.02268\times 10^{-2}$ & $-1.04495\times 10^{-4}$  \\\cline{2-8}
                              & $60 {}^{\circ}$& $6.83339\times 10^{-4}$   &$-3.32657\times 10^{-6}$  & $6.07821\times 10^{-3}$  & $-1.11064\times 10^{-4}$ & $4.32189\times 10^{-2}$ & $-1.50541\times 10^{-4}$  \\\cline{2-8}
                              & $80 {}^{\circ}$& $8.05842\times 10^{-4}$   &$-9.49684\times 10^{-7}$  & $3.6253\times 10^{-3}$  & $-1.90003\times 10^{-4}$ & $7.18506\times 10^{-2}$ & $-8.40648\times 10^{-5}$   \\\hline
                              & $20 {}^{\circ}$& $6.80194\times 10^{-4}$   &$-5.86914\times 10^{-6}$  & $8.62328\times 10^{-3}$  & $-7.76597\times 10^{-5}$ & $5.22018\times 10^{-3}$ & $-4.45666\times 10^{-5}$  \\\cline{2-8}
\multicolumn{1}{|c|}{$0.3$}   & $40 {}^{\circ}$& $7.26381\times 10^{-4}$   &$-5.84039\times 10^{-6}$  & $7.83838\times 10^{-3}$  & $-9.98056\times 10^{-5}$ & $2.04352\times 10^{-2}$ & $-1.44068\times 10^{-4}$  \\\cline{2-8}
                              & $60 {}^{\circ}$& $8.30438\times 10^{-4}$   &$-5.17799\times 10^{-6}$  & $6.49511\times 10^{-3}$  & $-1.6471\times 10^{-4}$ & $4.50611\times 10^{-2}$ & $-2.12317\times 10^{-4}$  \\\cline{2-8}
                              & $80 {}^{\circ}$& $1.08148\times 10^{-3}$   &$4.96873\times 10^{-9}$  & $4.36954\times 10^{-3}$  & $-3.44232\times 10^{-4}$ & $8.16276\times 10^{-2}$ & $-8.57416\times 10^{-5}$  \\\hline
                              & $20 {}^{\circ}$& $7.9204\times 10^{-4}$   &$-7.75248\times 10^{-6}$  & $8.28526\times 10^{-3}$  & $-9.01706\times 10^{-5}$ & $4.9126\times 10^{-3}$ & $-5.82767\times 10^{-5}$  \\\cline{2-8}
\multicolumn{1}{|c|}{$0.5$}   & $40 {}^{\circ}$& $8.65272\times 10^{-4}$   &$-8.07832\times 10^{-6}$  & $7.75244\times 10^{-3}$  & $-1.35035\times 10^{-4}$ & $1.97217\times 10^{-2}$ & $-1.92668\times 10^{-4}$  \\\cline{2-8}
                              & $60 {}^{\circ}$& $1.03918\times 10^{-3}$   &$-6.50404\times 10^{-6}$  & $6.83898\times 10^{-3}$  & $-2.63457\times 10^{-4}$ & $4.58815\times 10^{-2}$ & $-2.84301\times 10^{-4}$  \\\cline{2-8}
                              & $80 {}^{\circ}$& $4.63908\times 10^{-3}$   &$4.06096\times 10^{-5}$  & $1.5077\times 10^{-2}$  & $-1.12034\times 10^{-3}$ & $2.31186\times 10^{-1}$ & $1.64899\times 10^{-4}$  \\\hline

\end{tabular}
\end{table*}

\begin{figure}
    \centering
      \includegraphics[width=0.483\textwidth]{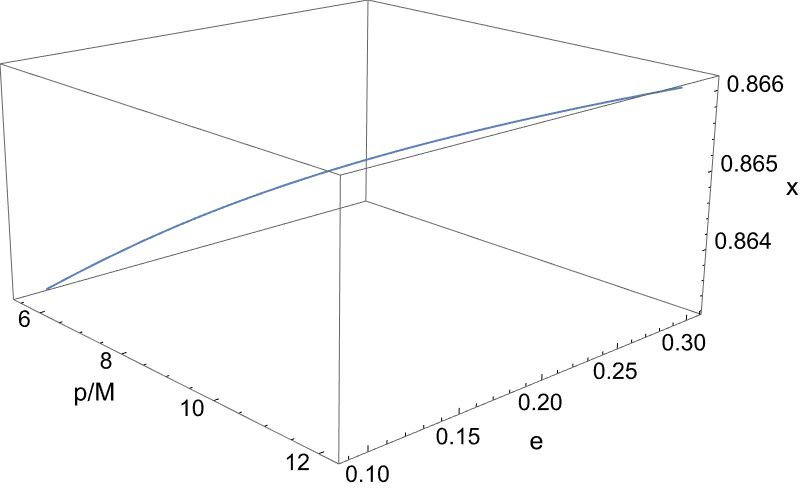}
    \caption{ Adiabatic evolution of the orbital parameters during an inspiral into Kerr black hole. The Kerr parameter is  $a=0.5M$ while the orbital parameters $(p,e,x)$ drift gradually from $(12M,0.3,0.8660)$ to $(6.08413, 0.10579, 0.86357)$.}  \label{fig:Inspiral}
\end{figure}

\begin{figure*}
     \centering
\subfloat{%
  \includegraphics[width=0.9\textwidth]{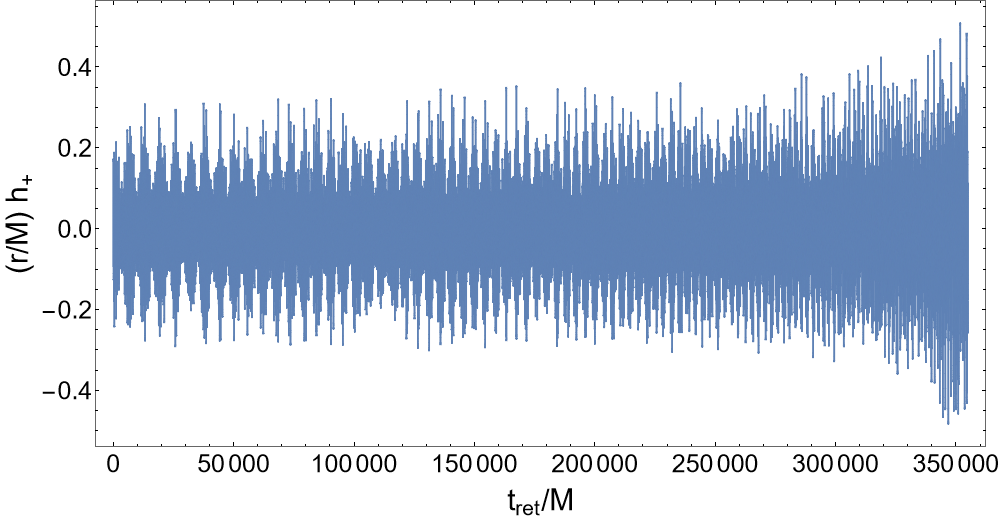}%
}  
  \hfill

 \subfloat{%
  \includegraphics[width=0.45\textwidth]{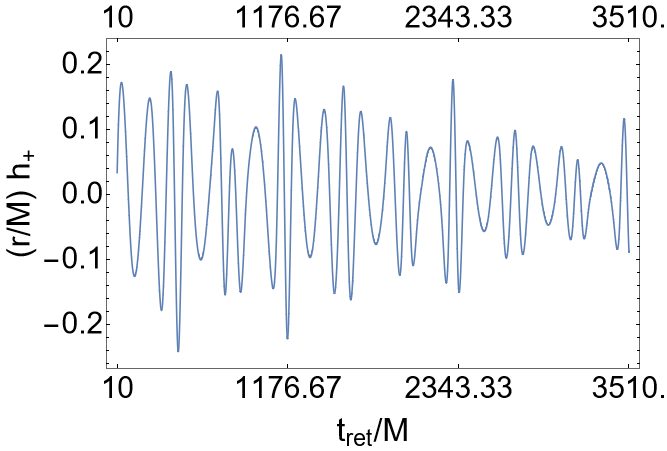}%
} 
 \subfloat{%
  \includegraphics[width=0.45\textwidth]{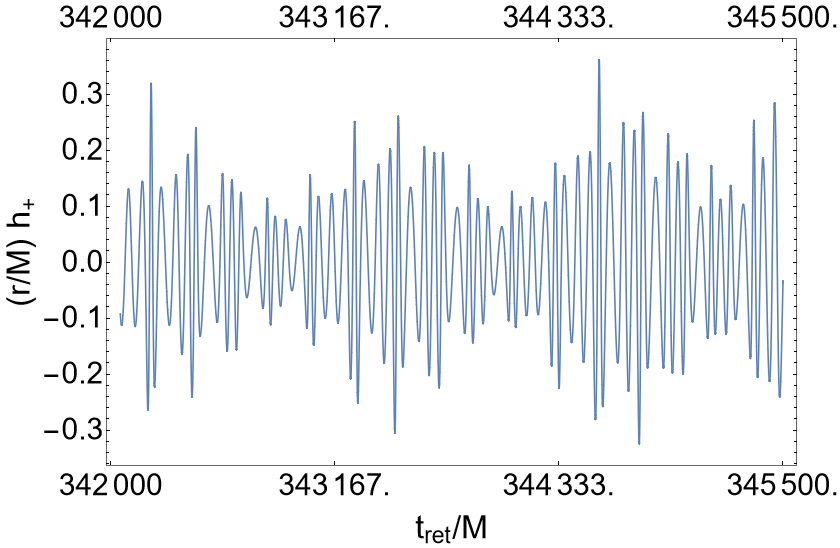}%
}      
        \caption{The component $h_{+}$ of the waveform produced during the inspiral shown in Fig.~\ref{fig:Inspiral} as observed from the equatorial plane. The complete waveform (top) is depicted in detail at the early stage (bottom left) and late  stage (bottom right) of the modelled inspiral. The mass ratio is $q=10^{-3}$.   }
        \label{fig:waveform}
\end{figure*}

\begin{figure*}[ht]
\begin{center}
 {\subfloat{\includegraphics[width=0.48\textwidth]{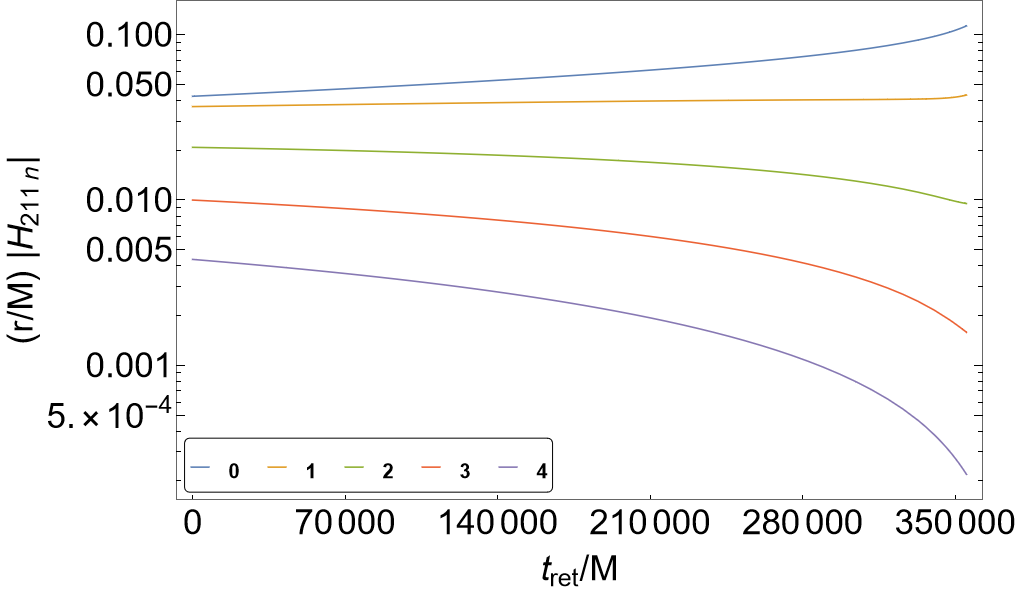}}
{  \subfloat{\includegraphics[width=0.48\textwidth]{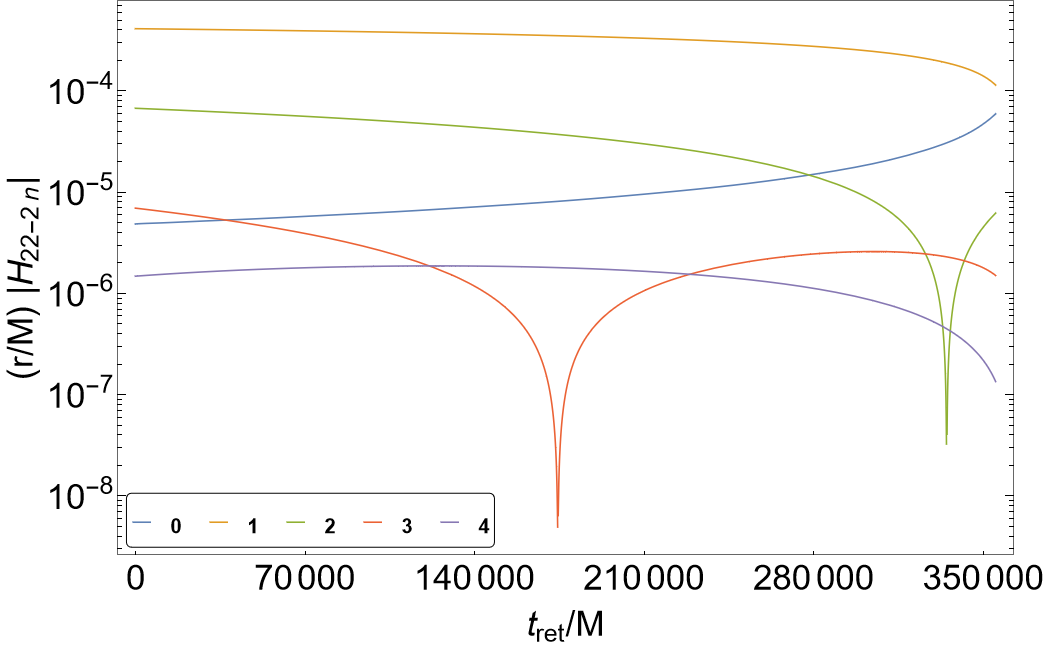}}}
}   
\end{center}
   \caption{Logarithmic plots of $H_{211n}$   (left) and $H_{22-2n}$ (right) during the inspiral shown in Fig.~\ref{fig:Inspiral}.}
        \label{fig:amplitudes}
\end{figure*}

We now include gravitational backreaction into our scheme. In this case, the secondary is inspiraling, and now not only the angles but also the actions change during the evolution. The equations describing the respective motion can then be expanded in mass ratio $q$ as  
\begin{align}
    \frac{\mathrm{d}\psi_i}{\mathrm{d}t}&= \Upsilon_i (\mathbf{J})+q\,f^{(1)}_i(\psi_r,\psi_u,\mathbf{J})+\mathcal{O}(q^2) ,  \\
    \frac{\mathrm{d} J_i}{\mathrm{d}t} &= q\, F^{(1)}_i(\psi_r,\psi_u,\mathbf{J})+\mathcal{O}(q^2),
\end{align} 
where $f^{(1)}_i$ provides the instantaneous first-order corrections to the geodesic frequency, while $F^{(1)}_i$ provides the flux of the actions (or equivalently of the orbital parameters).

Based on the two timescale separation, i.e. on the fact that the orbital timescale corresponding to the periods of the geodesic motion (determined by frequencies $ \Upsilon_i$) is much smaller than the inspiral timescale,  we can implement the adiabatic approximation. We average over the invariant torus parametrized by the angles in order to obtain averaged fluxes of the integrals of motion (actions) and hence the rate of change of the orbital parameters. The equations of motion at the leading order then read
\begin{align}
    \frac{\mathrm{d}\psi_i(t)}{\mathrm{d}t} &= \Upsilon_i (\mathbf{J(t)} ),
    \label{eq:adiabevol1}\\
        \frac{\mathrm{d} J_i(t)}{\mathrm{d}t} &= q\langle F^{(1)}_i\rangle (\mathbf{J(t)}),
    \label{eq:adiabevol2}
 \end{align}
where the averaged functions  $\langle F^{(1)}_i\rangle$ are the averaged fluxes.

To calculate the fluxes of $(p,e,\iota_0)$ we first have to calculate the fluxes of $\mathbf{C}=(E,L_z,Q)$. We employed the Teukolsky formalism to calculate the fluxes  of the three independent constants of motion $C_i$ through  the infinity and the primary's black hole horizon. This is in more detail described in Appendix ~\ref{sec:Teukolsky}. 
In full analogy with \cite{Drasco06} we present table~\ref{tab:fluxes} with the flux values $\Big\langle\frac{\mathrm{d}C_i^{\infty,H}}{\mathrm{d}t}\Big\rangle$ for $a=0.9M$, $p=6M$. 

It is difficult to determine the relative errors of the values in the table. It is clear, however, that the dominant contribution to the error comes from our approximation of the geodesic motion, this fact is more prominent for larger values of $e$. When discussing the results of the table~\ref{tab:fluxes} we have to keep in mind  that the values   $a=0.9M$, $p=6M$ are extreme in the sense that $0.9M$ is a large value for our perturbation  parameter $a$ while $p=6M$ means we are getting closer to the last stable spherical orbit. Hence, fluxes for larger values of $p$ and/or smaller values of $a$ would be more accurate. 

Regardless of the comment in the previous paragraph, we can conclude that the error grows with eccentricity which is an expected result since our geodesics  are derived using expansions from a Kerr spherical orbit and a Schwarzschild stable circular orbit. For $e=0.5$ the values of the fluxes cease to be reliable, since in most cases only the first significant digit seems to be correct (see the respective table in  \cite{Drasco06}.). An interesting aspect here is the dependence on the inclination that is non-trivial and it is relevant for high eccentricity and small semi-latus rectum, which can be seen particularly in the last row of the table~\ref{tab:fluxes}, where the error magnitude is the largest. This has its source already at the geodesic level, where we checked numerically that our approximation scheme does not behave well when the trajectory passes close to the central body.

 To better illustrate the error introduced by our geodesic approximation we created a table  of relative errors~\ref{tab:errors} corresponding to the Table~\ref{tab:fluxes}; the relative error is computed with respect to the fluxes from exact geodesics calculated through the {\em KerrGeodesics} package) and using the same new code for the Teukolsky amplitudes. As  already mentioned, the error depends on the eccentricity, for $e=0.1$ the relative error is sufficiently small and comparable with realistic values of mass ratio. On the other hand, the errors for  $e=0.5$ show the limits of the approximation. The values in the last row of the Table~\ref{tab:fluxes}  corresponding to the inclination $\iota_0=80 {}^{\circ}$ are of a particular interest   as the error is of the order of unity. This inclination dependence becomes noticeable only when close to the horizon. To be more specific, it is the radial part of the geodesic motion which deviates substantially. For instance, the relative error of the radial frequency is  $0.18$, while the angular part is well-behaved even for this highly inclined and eccentric orbit with error of the order $10^{-9}$ (see Table~\ref{tab:changeincl}.).  In general, we expect that such large errors in the fluxes should not consist a problem for initially not very eccentric orbits $e<0.5$, since by the time the inspiral would reach close to the horizon of the primary black hole the eccentricity will be sufficiently small. Details of how the errors in fluxes emerge are discussed in Appendix~\ref{sec:Teukolsky}.

\begin{table*}[]
 \caption{Table of relative errors of the total fluxes presented in table~\ref{tab:fluxes}. The relative errors are calculated with respect to the fluxes sourced by exact formulas for Kerr geodesics.} 
\centering
 \label{tab:errors}
\begin{tabular}{|l|c|c|c|c|c|c|c|}
\hline
\multicolumn{1}{|c|}{$e$}     & $\iota_0$&$\delta\Big\langle\frac{\mathrm{d}E^{\infty}}{\mathrm{d}t}\Big\rangle= \delta \dot{E}^{\infty} $&$ \delta \dot{E}^{H}$ &$ \delta \dot{L}_z^{\infty}$ & $\delta \dot{L}_z^{H}$   &$\delta \dot{Q}^{\infty}$ &$\delta \dot{Q}^{H}$\\ \hline
                              & $20 {}^{\circ}$& $4.27\times 10^{-7}$   &$1.12\times 10^{-8}$  & $3.88\times 10^{-7}$  & $4.45\times 10^{-8}$ & $5.05\times 10^{-8}$ & $2.18\times 10^{-7}$  \\\cline{2-8}
\multicolumn{1}{|c|}{$0.1$}   & $40 {}^{\circ}$& $6.17\times 10^{-7}$   &$1.10\times 10^{-7}$  & $5.39\times 10^{-7}$  & $1.51\times 10^{-7}$ & $4.01\times 10^{-7}$ & $1.88\times 10^{-7}$  \\\cline{2-8}
                              & $60 {}^{\circ}$& $6.90\times 10^{-8}$   &$2.99\times 10^{-8}$  & $8.35\times 10^{-8}$  & $2.61\times 10^{-8}$ & $6.77\times 10^{-8}$ & $1.78\times 10^{-7}$  \\\cline{2-8}
                              & $80 {}^{\circ}$& $5.54\times 10^{-7}$   &$2.88\times 10^{-7}$  & $4.82\times 10^{-7}$  & $2.56\times 10^{-7}$ & $3.37\times 10^{-7}$ & $1.02\times 10^{-6}$   \\\hline
                              & $20 {}^{\circ}$& $1.84\times 10^{-4}$   &$9.51\times 10^{-5}$  & $1.85\times 10^{-4}$  & $5.41\times 10^{-6}$ & $1.57\times 10^{-4}$ & $1.84\times 10^{-4}$  \\\cline{2-8}
\multicolumn{1}{|c|}{$0.3$}   & $40 {}^{\circ}$& $1.45\times 10^{-4}$   &$1.78\times 10^{-4}$  & $1.51\times 10^{-4}$  & $7.07\times 10^{-4}$ & $1.22\times 10^{-4}$ & $1.54\times 10^{-4}$  \\\cline{2-8}
                              & $60 {}^{\circ}$& $1.13\times 10^{-4}$   &$2.95\times 10^{-4}$  & $1.28\times 10^{-4}$  & $3.41\times 10^{-4}$ & $8.98\times 10^{-5}$ & $1.49\times 10^{-4}$  \\\cline{2-8}
                              & $80 {}^{\circ}$& $1.60\times 10^{-3}$   &$4.32\times 10^{-1}$  & $1.39\times 10^{-3}$  & $1.02\times 10^{-3}$ & $1.15\times 10^{-3}$ & $6.19\times 10^{-4}$  \\\hline
                              & $20 {}^{\circ}$& $8.17\times 10^{-3}$   &$7.09\times 10^{-2}$  & $7.06\times 10^{-3}$  & $1.29\times 10^{-2}$ & $6.66\times 10^{-3}$ & $9.77\times 10^{-3}$  \\\cline{2-8}
\multicolumn{1}{|c|}{$0.5$}   & $40 {}^{\circ}$& $1.01\times 10^{-2}$   &$9.61\times 10^{-2}$  & $8.30\times 10^{-3}$  & $1.25\times 10^{-2}$ & $8.25\times 10^{-3}$ & $9.50\times 10^{-3}$  \\\cline{2-8}
                              & $60 {}^{\circ}$& $1.83\times 10^{-2}$   &$1.22\times 10^{-1}$  & $1.56\times 10^{-2}$  & $1.94\times 10^{-2}$ & $1.43\times 10^{-2}$ & $1.14\times 10^{-2}$  \\\cline{2-8}
                              & $80 {}^{\circ}$& $1.77\times 10^{0}$   &$1.40\times 10^{-2}$  & $1.56\times 10^{0}$  & $2.19\times 10^{-1}$ & $1.28\times 10^{0}$ & $3.84\times 10^{0}$  \\\hline

\end{tabular}
\end{table*}

Once the fluxes of constants of motion are calculated we can easily determine the rate of change of the three orbital parameters $(p,e,x)$ where $x=\cos(\iota_0)$. For this, we have created a grid in the $(p,e,x)$ space to interpolate the fluxes (see Appendix~\ref{sec:Teukolsky}). Solving the system of equations~\eqref{eq:adiabevol2} is then a straightforward task with the result being the three parameters as functions of the coordinate (Boyer-Lindquist) time $(p(t),e(t),x(t))$.

We are now going to demonstrate our scheme on an inspiral with $a=0.5M$ and initial condition 
\begin{align*}
    (p_{\rm ini},e_{\rm ini},x_{\rm ini})= (12M,0.3,\cos(30 {}^{\circ}) )=(12M,0.3,0.86602) .
\end{align*}
After adiabatically evolving the parameters we get   
\begin{align*}
    (p_{\rm fin},e_{\rm fin},x_{\rm fin})=(6.08413, 0.10579, 0.86357),
\end{align*}
at the coordinate time $t_{\rm insp}=355M /q$. The evolution of this eccentric non-equatorial inspiral is depicted in Fig.~\ref{fig:Inspiral}. Note that the eccentricity of the inspiral becomes sufficiently small ($e\sim 0.1$) by the end of the computation, i.e. by the time it gets closer to the horizon.

Inspired by Ref.~\cite{Hughes21}, we calculate the gravitational waveforms. The strain $h=h_{+}+i h_{\times}$ can for $r\rightarrow \infty$ be written as a  sum of individual modes
\begin{align}
    h =\frac{1}{r} \sum_{lmkn} H_{lmkn}&(t_{\rm ret},\theta) e^{-i\Phi_{mkn}(t_{\rm ret}) + i m \phi}, \label{eq:sum}\\
    H_{lmkn}(t_{\rm ret},\theta)=-2& \frac{C^+_{lmkn}(t_{\rm ret})}{\omega_{mkn}^2(t_{\rm ret})} {}_{-2}S_{lm}(\theta,a \omega_{mkn}(t_{\rm ret})),
\label{eq:wave}
\end{align}    
where  $t_{\rm ret}$ is the retarded time which can be at the infinity written as $t_{\rm ret}\approx t-r$. The amplitudes $H_{lmkn}$ can be expressed in terms of the coefficients $C^+_{lmkn}$, which are calculated by solving the inhomogeneous Teukolsky equation (see Appendix ~\ref{sec:Teukolsky}), while ${}_{-2}S_{lm}$ are the spin-weighted spheroidal harmonic functions. 

The frequencies $ \omega_{mkn}$ and the phases  $\Phi_{mkn}$ read
\begin{align}
  \omega_{mkn}=& n\Omega_r+k\Omega_u+m \Omega_\phi,\\
\Phi_{mkn}(t_{\rm ret})=&\displaystyle\int_{t_{ret0}}^{t_{\rm ret}} \omega_{mkn}(t) \mathrm{d}t,
    \label{eq:Phases}
\end{align}
where  $\Omega_i=\Upsilon_i/\Upsilon$ are the fundamental frequencies with respect to the time $t$, which are calculated using Eqs.~\eqref{eq:rfrec}-\eqref{eq:tfrec}.

Knowing the trajectory in the parameter space $(p(t),e(t),x(t))$, we can adiabatically evolve the frequencies $\Omega_i$ and  $C^+_{lmkn}$, since they are functions of the orbital parameters. The integration of frequencies in Eq.~\eqref{eq:Phases} is just a consequence of the adiabatic evolution of angles in Eq.~\eqref{eq:adiabevol1}.

Before plotting a waveform we have to fix the mass ratio $q$ because in Eq.~\eqref{eq:sum} both timescales are present, the rapidly oscillating phases $\Phi_{mkn}$ and the slowly changing amplitudes $H_{lmkn}$. In our case, we calculated the waveform from the generic inspiral provided in Fig.~\ref{fig:Inspiral}) and chose the mass ratio to be $q=10^{-3}$, which is not realistic for an EMRI, but shortens the time of the calculation of the  complete waveform whose component  $h_{+}$ is shown in Fig.~\ref{fig:waveform} for illustration. Similarly to \cite{Hughes21}, we  plot the evolution of the amplitudes for given $l$, $m$, $k$ and $n$ to see how their contribution to the sum~\eqref{eq:sum} changes over the time of the inspiral (Fig.~\ref{fig:amplitudes}).

\section{Conclusion} \label{sec:Conc}

Writing a Hamiltonian system in action-angle variables provides characteristic quantities of the system, like the fundamental frequencies, in closed form. The split in actions (constants of motion) and angles is a very convenient approach for modelling EMRIs, when adopting the two time-scale approximation. Having this in mind, we have used the Lie series transformation method to express the Hamiltonian giving the geodesic equation of motion in a Kerr spacetime in action angle variables. The advantage in this approach is that all the involved relations are in closed form and the transformation is invertible. Hence, one can go back and forth easily between the Boyer-Lindquist coordinates and the action-angle variables. The drawback is that the approach is perturbative and the Hamiltonian in action-angle variables only approximates the original Hamiltonian. However, one should keep in mind that every model is just an approximation of the real system and what is really required from a model is to be accurate enough for the purpose we need it to serve.

Taking advantage of the fact that the Mino-Carter time allows us to have a separable Hamiltonian for the original system, we split it into a radial part and an angular part. The Hamiltonian function can be expressed purely in the constants of motion and the radial coordinate along with its conjugate momentum, while the Carter constant can be expressed purely in the constants of motion $E,~L_z$ and the polar coordinate along with its conjugate momentum. This allowed us to use different perturbation schemes for each part. Namely, for the radial part, we perturbed around a spherical geodesic orbit in Kerr spacetime, while for the angular we perturbed around an inclined geodesic circular orbit in Schwarzschild spacetime. The latter choice was inspired by the fact that for Schwarschild the inclination is constant and the introduction of the Kerr parameter essentially causes an oscillation around this plane. Actually, this is the reason we preferred to transform first the system from the Boyer-Lindquist coordinates to the polar nodal one before we started the perturbation procedure.

By applying the Lie series method we noticed that after a certain number of transformations, the actions ceased to converge fast to a constant value. This convergence saturation defined the number of transformations we employed to define our Hamiltonian function $H_{AA}$ expressed purely in actions, which approximates the original system.  We found that the approximation can be considered satisfactory for eccentricities $e\leq 0.5$; as expected the lower the eccentricity the better the approximation. We found also a weak dependency of the accuracy on the value of the inclination and the value of the Kerr parameter, but these dependencies are insignificant in comparison to the effect that eccentricity has on our approximation.    

Having transformed the system in action-angle variables, we used it to model the adiabatic evolution of an EMRI as a showcase of what is possible. The first step for this was the calculation of the fluxes of the constants of motion at infinity and through the horizon of the primary black hole. We used a newly developed frequency domain solver. By comparing our approximative fluxes values with those given in \cite{Drasco06} and those computed using the exact solutions of the geodesic orbits in Kerr, we found them to be in good agreement in the domain that our approximation is valid. However, the accuracy in the fluxes is significantly lower than the one reached on the level of the actions. In order to achieve better accuracy, further improvement in our approximative scheme of Kerr geodesics is needed. After this test, we provided  an example of adiabatic evolution on a generic inspiral in the Kerr background.

Providing the Hamiltonian of geodesic motion in Kerr in actions and having the whole transformation in closed form allows for several useful applications apart from being able to evolve efficiently an EMRI in the adiabatic approximation. Namely, one can include several forms of perturbations to an EMRI system, like matter distribution around the primary black hole or another stellar compact object in the vicinity of the EMRI, and slightly extend the provided scheme, as was done in \cite{Polcar22} for the Schwarzschild background, to be able to evolve and study EMRIs in perturbed systems as well.

\section*{Acknowledgements}

MK, LP,  VS and GLG have been supported by the fellowship Lumina Quaeruntur No. LQ100032102 of the Czech Academy of Sciences. L.P. and V.S. acknowledge support by the project "Grant schemes at CU" (reg.no. CZ.02.2.69/0.0/0.0/19\_073/0016935). Computational resources were supplied by the project "e-Infrastruktura CZ" (e-INFRA CZ LM2018140) supported by the Ministry of Education, Youth and Sports of the Czech Republic. We would like to thank Vojt\v{e}ch Witzany for his comments on our work.

\bibliographystyle{unsrt}
\bibliography{refs}

\appendix
\section{Polar-nodal coordinate}\label{sec:polnod}

To describe the generic motion of a particle in a central force field a Cartesian coordinate system $\boldsymbol{x}=\lbrace x, y,z  \rbrace$  originating at the position of the force centre is usually not the best choice. A coordinate system in which the coordinates and their conjugate moments can be related to the orbital parameter of the motion sounds as a better idea. Therefore, we are interested in the polar-nodal coordinate system and the Euler angles~\cite{palacian,hill,lara}.

Let us assume a set of Cartesian coordinates $\boldsymbol{x}=\lbrace x, y,z  \rbrace$ along with their conjugate momenta $\boldsymbol{p}=\lbrace p_x,p_y,p_z\rbrace$. In this set of variables, the total angular momentum vector is defined as $\boldsymbol{p_u}= \boldsymbol{x} \times \boldsymbol{p}$. We can decompose this vector as $\boldsymbol{p_u}= p_u \boldsymbol{n}$, where $p_u > 0$ is the measure of the angular momentum and the unit vector $\boldsymbol{n}$, i.e. $\mid\mid \boldsymbol{n}\mid\mid=1$, is perpendicular to the instantaneous orbital plane.

The inclination of the orbital plane denoted by $\iota$, is determined from the angle between the orbital plane and the equatorial plane or $\boldsymbol{z}\cdot \boldsymbol{n}= \cos \iota$. The angle $\nu$ between the positive $x$ axis and the lines of nodes is called the longitude of the ascending node. The line of nodes is the intersection of the orbital plane and the equatorial plane, while the nodes are those two points where the particle passes the equatorial plane. The ascending node is the node where the particle passes the equatorial plane from $-z$ to $+z$. The line of nodes lies along a vector $\boldsymbol{l}$ such that $\boldsymbol{z}\times \boldsymbol{n}= \boldsymbol{l} \sin \iota$. Finally, on the orbital plane, the angle between the ascending node and the particle is defined as the argument of the latitude $u$. The angles $\lbrace\nu, u, \iota \rbrace$ are known as the Euler angles and the three unit vectors $\lbrace \boldsymbol{n}, \boldsymbol{l}, \boldsymbol{n} \times \boldsymbol{l}\rbrace$ define the nodal frame.  Note that, both unit vectors $\boldsymbol{l}$ and $\boldsymbol{n}\times \boldsymbol{l}$ lie in the orbital plane.

 \begin{figure}
    \centering
      \includegraphics[width=0.483\textwidth]{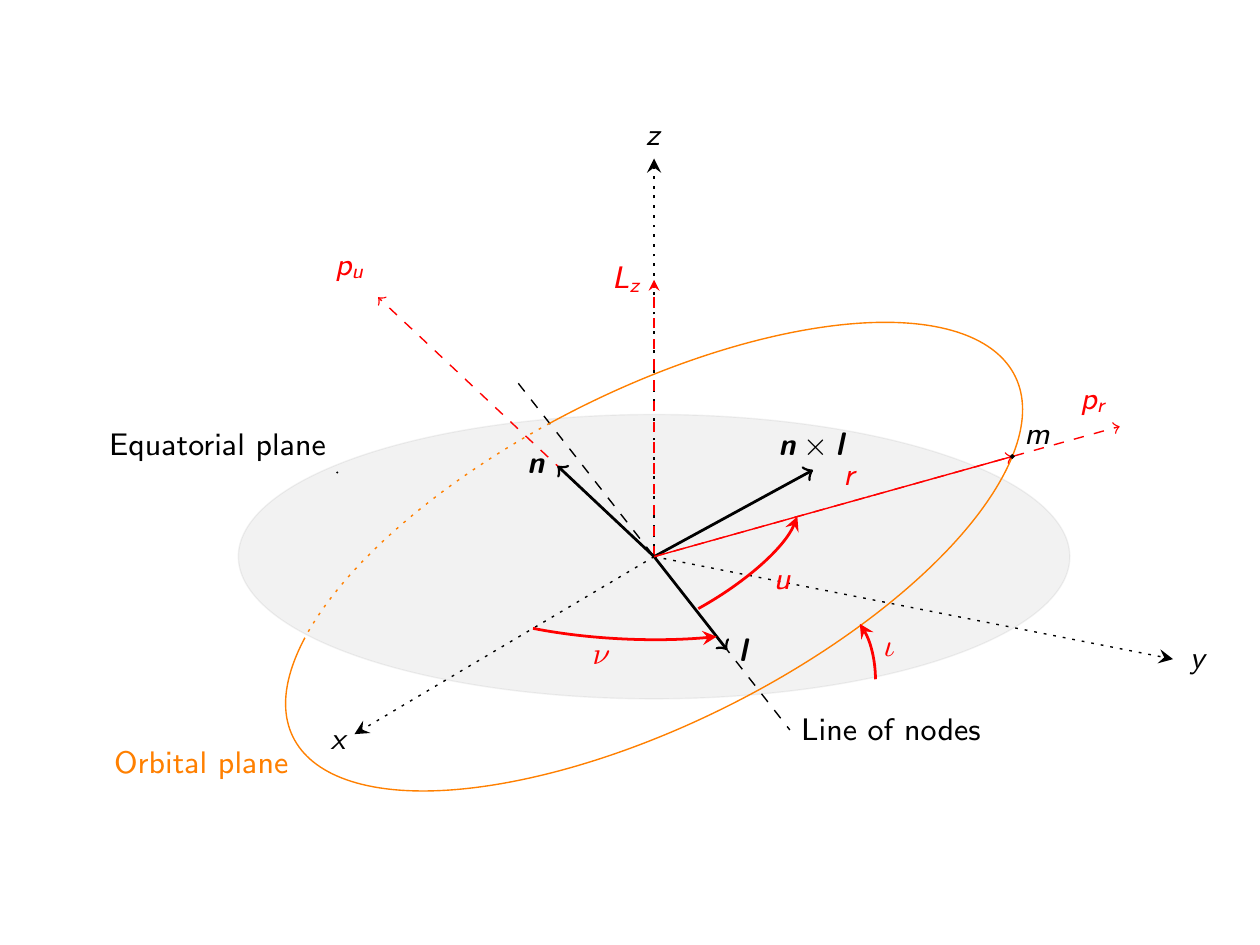}
    \caption{ This figure shows the Euler angles $\lbrace\nu, u, \iota \rbrace$, the nodal frame $\lbrace \boldsymbol{n}, \boldsymbol{l}, \boldsymbol{n} \times \boldsymbol{l}\rbrace$, and the polar-nodal variables $\lbrace r, u, \nu , p_r, p_u, L_z\rbrace$. The unit vector $\boldsymbol{n}$ is perpendicular to the instantaneous orbital plane while the unit vector $\boldsymbol{l}$ lies in the intersection of the equatorial plane and the orbital plane. From $\boldsymbol{n}$ and $\boldsymbol{l}$ the unit vector $\boldsymbol{n}\times \boldsymbol{l}$, which lies on the orbital plane, is defined. The motion of the orbiting body can be described by its radial distance from the central body $r$, the argument of the latitude $u$, and the longitude of the ascending node $\nu$ along with their conjugate moments $p_r$, $p_u$, and $L_z$ respectively. } \label{fig:polnodcoord}
\end{figure}

The polar-nodal coordinate consists of the coordinates $\lbrace r, u, \nu \rbrace$ and their conjugate momenta $\lbrace p_r, p_u, L_z\rbrace$. Fig~\ref{fig:polnodcoord} illustrates these variables. The transformation between Cartesian and polar nodal coordinates are given by~\cite{hill}
\begin{align}\label{eq:cartespl}
    x&= r \cos u \cos \nu - r \sin u \cos \iota \sin \nu,\nonumber\\
    y&= r \cos u \sin \nu + r \sin u \cos \iota \cos \nu,\nonumber\\
    z&= r \sin u \sin \iota,\nonumber\\
    p_x&= p_x^\prime \cos\nu - p_y^\prime \cos\iota \sin\nu,\nonumber\\
    p_y&= p_x^\prime \sin\nu + p_y^\prime \cos\iota \cos\nu,\nonumber\\
    p_z&= p_y^\prime \sin \iota,
\end{align}
where
\begin{align}
    p_x^\prime&= p_r \cos u- \frac{p_u}{r} \sin u,\nonumber\\
    p_y^\prime&= p_r \sin u+ \frac{p_u}{r} \cos u\\
    \cos \iota &= L_z/ p_u.\nonumber
\end{align}

Inverting the transformation~\eqref{eq:cartespl} results in
\begin{align}\label{eq:spherpl}
    r&= \sqrt{x^2+y^2+z^2},\nonumber\\
    \theta&= \arccos\frac{z}{r}= \arccos \left ( \sqrt{1-\frac{L_z^2}{p_u^2}} \sin u \right),\nonumber\\
    \phi&= \arctan \frac{x}{y}=  u+\nu + \arctan\left(\frac{(L_z/p_u-1) \cos u \sin u}{1+ (L_z/p_u-1) \sin^2 u} \right),
\end{align}
and also
\begin{align}\label{eq:sphermomentpl}
    p_r&= \frac{1}{r} \boldsymbol{x} \cdot \boldsymbol{p},\nonumber\\
    p_u&= \sqrt{(\boldsymbol{x}\times \boldsymbol{p})^2},\nonumber\\
    p_\theta&= \sqrt{p_u^2-\frac{L_z^2}{\sin \theta^2}}=\pm p_u \sqrt{1-\frac{L_z^2}{p_u^2-(p_u^2-L_z^2) \sin^2 u}},\nonumber\\
    p_\phi&= L_z= x p_y- y p_x= p_u \cos \iota.
\end{align}
From $p_\phi$ it's obvious that  $\cos \iota= L_z/ p_u$, in other words, the inclination angle $\iota$ is not an independent variable since it can be written in terms of $p_u$ and $L_z$. This is why the inclination angle does not appear in the polar-nodal coordinates.

\section{\textit{CPKerrGeodesics} Package}

This section provides some information regarding the \textit{CPKerrGeodesics} package \cite{CPKerrGeodesics}, i.e.\textit{ Canonically Perturbed Kerr Geodesics}. 

This  package provides the Hamiltonian, orbital frequencies, and trajectories in the AA variables once the
parameters $\lbrace a, p, e, \iota_0 \rbrace$, i.e. Kerr parameter, semilatus
rectum, eccentricity and initial inclination respectively,
are provided. 

As we discussed in Sec.~\ref{sec:NewHam}, we set $n=10$ and $n^\prime = 7$, namely $10$ canonical transformations for the radial part and $7$ canonical transformations for the angular part to provide reliable orbital results for eccentricities smaller than $0.5$. Then, we derive the explicit formula for the Hamiltonian~\eqref{eq:Haa} which is a function of $\lbrace J_r, J_u, J_{\nu}, J_t \rbrace $. Consequently, the frequencies are determined from Eqs.~\eqref{eq:rfrec}-~\eqref{eq:tfrec}. The trajectories are determined from Eqs.~\eqref{eq:rnew}-~\eqref{eq:tnew} which are the functions of actions and angles, i.e. $\lbrace J_r, J_u, J_{\nu}, J_t \rbrace $ and $\lbrace \psi_r, \psi_u, \psi_\nu, \psi_t \rbrace$. 

In order to determine the trajectories~\eqref{eq:rnew}-~\eqref{eq:tnew} in terms of the Mino-time $\lambda$ then it does the following steps
\begin{itemize}
    \item  From the given input, i.e. $\lbrace a, p, e, \iota_{0} \rbrace$, the code determines numerically the actions $\lbrace J_r, J_u, J_{\nu}, J_t \rbrace$ from Eqs.~\eqref{eq:jrpl}-~\eqref{eq:jupl} and $L_z$ and $E$ from~\cite{Schmidt02}.
    \item  The numerical values of actions then determine the frequencies and consequently $\psi_i (\lambda)= \Upsilon_i \lambda$.
\end{itemize}

Thus, substituting these two steps into the Eqs.~\eqref{eq:rnew}-~\eqref{eq:tnew} determines the trajectories as a function of $\lambda$. 

This package was inspired by  {\it KerrGeodesics} package from the  \textit{Black Hole Perturbation Toolkit} \cite{BHPT}. The part of our package which calculates the constants of motion from the   parameters  $\lbrace a, p, e, \iota_0 \rbrace$ was directly adopted from the  {\it KerrGeodesics} package.

\section{Two canonical transformations}\label{sec:2ct}

In this section we provide an example of a Hamiltonian function expressed in AA variables after two canonical transformations are performed; the respective generating functions are provided as well. We also provide the trajectories in the AA variables when we applied one canonical transformation. In these relations, many constants appear; some of them are given in Eqs.~\eqref{eq:Ec},~\eqref{eq:Qc}, and~\eqref{eq:fisrtconst}, and the others are given in Sec.~\ref{sec:consts}.

\paragraph{\textbf{Initial conditions.}} For a given parameter set  $\lbrace a, p, e, \iota_0\rbrace$, we set the radius of the circular orbit $$r_c=\frac{p}{1-e^2}+e (1-10\,e) M$$ and $\delta= r_c\, a/ (e M^2)$ as we mentioned in Sec.~\ref{sec:NewHam}. The $L_{zc}$ is chosen in such a way that the relation  $L_{zc}= \cot \iota_0 \sqrt{Q_c}$ satisfies Eq.~\eqref{eq:Qc}; and we set $\Tilde{L}_{zc}= \cos \iota_0\, p_{uc}$. 

\paragraph{\textbf{The Hamiltonian in AA variables.}}\label{sec:ham2} The Hamiltonian~\eqref{eq:Haa} for $n=2$ and $n^\prime=2$ has the following form 
\begin{align} \label{eq:ham2}
    H&=\left(\Omega_{t0} J_t  + \frac{1}{2} \Tilde{Q}  +\Omega_{r0} J_r  +\Omega_{z0} J_{\nu}  \right)- \nonumber \\
     &\Big(\frac{3 J_r^2 (A^2_{rs}+A^2_{s3})+ J_r A_{rs} A_{tzs}(J_t,J_{\nu})+ A^2_{tzs}(J_t,J_{\nu})}{\Omega_{r0}} \nonumber \\
     &- J_r^2 B_r- J_r B_{rtz}(J_t,J_{\nu})- B_{tz}(J_t,J_{\nu}) \Big),
\end{align}
where 
\begin{align}\label{eq:carter2}
        \Tilde{Q}&=2 \left(p_{uc} J_u- \Tilde{L}_{zc} ( J_{\nu}+ L_{zc}- \Tilde{L}_{zc})\right)-\big( ( J_{\nu}+ L_{zc}- \Tilde{L}_{zc})^2\nonumber\\
    &-J_u^2\big)+\frac{a^2}{ p_{uc}^3}\big((J_t+ p_{tc}-\Tilde{p}_{tc}) \Tilde{p}_{tc} p_{uc}\left(\Tilde{L}_{zc}^2-p_{uc}^2\right)\nonumber\\
  & +(J_{\nu}+ L_{zc}- \Tilde{L}_{zc}) \Tilde{L}_{zc} \left(\Tilde{p}_{tc}^2-1\right) p_{uc} \nonumber\\
  & -J_u \Tilde{L}_{zc}^2 \left(\Tilde{p}_{tc}^2-1\right)\big)-Q_c,
\end{align}
and 
\begin{align}
    A_{tzs}(J_t,J_\nu)&=\frac{4 \beta}{\alpha}^{1/4} ( J_t   b_t+J_{\nu} \,b_z),\label{eq:atzs}\\
    B_{rtz}(J_t,J_\nu)&= \sqrt{\frac{\beta}{\alpha}} ( c_z J_{\nu}+ c_t J_t),\label{eq:brtz}\\
B_{tz}(J_t,J_\nu)&= c_{t^2}\, J_t^2+ c_{tz} \,J_t\, J_{\nu}+ c_{z^2} \,J_{\nu}^2.\label{eq:btz}
\end{align}
Substituting Eqs.~\eqref{eq:carter2}-~\eqref{eq:btz} into the Hamiltonian~\eqref{eq:ham2} provides the approximate Hamiltonian in actions. 

\paragraph{\textbf{Generating functions.}}\label{sec.gener2}
The radial generating functions for this system read
\begin{align}
\chi_{r1}&=-\epsilon \frac{\sqrt{J_r}}{3 \Omega_{r0}} \left(3 (A_{rs} J_r + A_{tzs}) \cos \psi_r+ A_{s3} J_r \cos(3 \psi_r)  \right),\nonumber \\
\chi_{r2}&=\epsilon^2 \frac{J_r}{8 \Omega_{r0}}\Big[2 \Omega_{r0}(2 ( B_{rc2} J_r- B_{rtz}) \sin( 2 \psi_r) \nonumber\\
&+ B_{rc4} J_r \sin( 4 \psi_r))-( 8 A_{rs} A_{s3} J_r+4 A_{s3} A_{tzs}) \sin(2 \psi_r)\nonumber\\
&- A_{rs} A_{s3} J_r \sin(4 \psi_r)\Big],
\end{align}
and the angular ones are 
\begin{align}
    \chi_{u1}&=\frac{a^2 \epsilon^2 \left(\Tilde{p}_{tc}^2-1\right) \left(p_{uc}^2-\Tilde{L}_{zc}^2\right) \sin (2 \psi_u)}{8 p_{uc}^3},\nonumber\\
    \chi_{u2}&=\frac{a^2 \epsilon^4 \sin (2 \psi_u)}{8 p_{uc}^4}\Big[2\,(J_t+ p_{tc}-\Tilde{p}_{tc}) \,\Tilde{p}_{tc} \,p_{uc}^3\nonumber\\
    &+J_u \left(\Tilde{p}_{tc}^2-1\right) \left(3 \Tilde{L}_{zc}^2-p_{uc}^2\right)\nonumber\\
    &-2 \Tilde{L}_{zc} \,p_{uc} \big((J_t+ p_{tc}-\Tilde{p}_{tc}) \Tilde{L}_{zc} \Tilde{p}_{tc} \nonumber\\
&+( J_{\nu}+ L_{zc}- \Tilde{L}_{zc}) \big(\Tilde{p}_{tc}^2-1\big)\big)\Big].
\end{align}

\paragraph{\textbf{Trajectories.}}\label{sec:traj2}
The trajectories in terms of the AA variables determine from Eqs.~\eqref{eq:rnew}-~\eqref{eq:tnew}. Here we apply only one canonical transformation\footnote{Only for $t_{\rm new}$ we apply two canonical transformations, i.e. $\chi_{u_1}$ and $\chi_{u_2}$ since $\chi_{u_1}$ does not depend on $J_t$ and does not have any effect on it.}, i.e. $\chi_{r_{1}}$ and $\chi_{u_1}$, to derive the following trajectories
\begin{widetext} 
\begin{align} 
\begin{split}
    r_{\rm new}&=\left(r_c+ \delta \sqrt{\frac{2 J_r}{m_c \Omega_{r0}}} \sin (\psi_r)\right)-\frac{  \delta}{2 \alpha \sqrt{\alpha  \beta}}
    \Big[\sqrt{\alpha  \beta } (b_t J_t+b_z J_z)+ J_r (b_{p^2} \alpha+3 b_{r^2} \beta)+J_r (b_{p^2} \alpha+ b_{r^2} \beta) \cos (2 \psi_r)\Big],\\
    \theta_{\rm new}&=\arccos \left(\sqrt{1-\frac{(J_{\nu}+L_{zc})^2}{(J_{u}+p_{uc})^2}} \sin \psi_u\right)+
    \frac{a^2(J_{\nu}+L_{zc})^2\left(\Tilde{p}_{tc}^2-1\right) \left(\Tilde{L}_{zc}^2-p_{uc}^2\right) (\sin (\psi_{u})-\sin (3 \psi_{u}))}{8 p_{uc}^3\,(J_{u}+p_{uc})^3 \sqrt{1-\frac{(J_{\nu}+L_{zc})^2}{(J_{u}+p_{uc})^2}} \sqrt{\sin ^2(\psi_u) \left(\frac{(J_{\nu}+L_{zc})^2}{(J_{u}+p_{uc})^2}-1\right)+1}},\\
    \phi_{\rm new}&=\psi_\nu+\psi_u- \Bigg[\frac{b_z \sqrt{J_r} \cos{\psi_r}}{\sqrt{2 \alpha \sqrt{\alpha \beta}}}\Bigg]
    + \Bigg[\arctan\left(\frac{\left(\frac{J_{\nu}+L_{zc}}{J_{u}+p_{uc}}-1\right) \cos \psi_u \sin \psi_u}{1+ \left(\frac{J_{\nu}+L_{zc}}{J_{u}+p_{uc}}-1\right) \sin^2 \psi_u} \right)+\nonumber\\
    &+\frac{a^2 (J_{\nu}+L_{zc}) \left(\Tilde{p}_{tc}^2-1\right) \left(p_{uc}^2-\Tilde{L}_{zc}^2\right) \sin (4 \psi_u)}{8 p_{uc}^3 \left(\cos (2 \psi_u) \left(J_u^2+2 J_u p_{uc}-(J_{\nu}+L_{zc})^2+p_{uc}^2\right)+J_u^2+2 J_u p_{uc}+(J_{\nu}+L_{zc})^2+p_{uc}^2\right)}\Bigg],\\
    t_{\rm new}&=\psi_t- \Bigg[\frac{b_t \sqrt{J_r} \cos{\psi_r}}{\sqrt{2 \alpha \sqrt{\alpha \beta}}}\Bigg]+\Bigg[\frac{a^2  \Tilde{p}_{tc} \sin (2 \psi_u)}{4 p_{uc}}-\frac{a^2  \Tilde{L}_{zc}^2 \Tilde{p}_{tc} \sin (2 \psi_u)}{4 p_{uc}^3} \Bigg].
\end{split}
\end{align}
\end{widetext}

\paragraph{\textbf{Constants.}}\label{sec:consts}
The constants which appeared in the Secs.~\ref{sec:ham2},~\ref{sec.gener2}, and~\ref{sec:traj2} are given by 
\begin{align} 
A_{rs}&= \sqrt{\frac{\alpha^{1/2} \beta^{3/2}}{2}} \left(\frac{b_{p^2}}{\beta}+3 \frac{b_{r^2}}{\alpha} \right),\\
A_{s3}&= \sqrt{\frac{\alpha^{1/2} \beta^{3/2}}{2}} \left(\frac{b_{p^2}}{\beta}- \frac{b_{r^2}}{\alpha} \right),
\end{align}
\begin{align}
    B_r&=\frac{3\,\beta \, c_{r^2}}{2 \, \alpha}+\frac{1}{4},\\
B_{rc2}&= -\frac{2 \, \beta\, c_{r^2}}{\alpha},\\
B_{rc4}&=\frac{\,\beta \, c_{r^2}}{2 \, \alpha}-\frac{1}{4},
\end{align}
where
\begin{align}
b_{r^2}&= \frac{2 \delta^3}{\Delta_c^4}\left( \mathfrak{B}_{1}\,p_{tc}^2+\mathfrak{B}_{2}\, p_{tc} L_{zc}+ \mathfrak{B}_{3}\, L_{zc}^2\right)\\
b_t&=-\frac{2 \delta}{\Delta_c^3}\left( \mathfrak{B}_4 p_{tc} +\mathfrak{B}_5 L_{zc}\right),\\
b_z&=\frac{2 a \delta }{\Delta_c^2}\left(M \left(r_c^2-a^2\right) p_{tc} +a (r_c-M) L_{zc}\right),\\
b_{p^2}&=\frac{(r_c-M)}{\delta },
\end{align}
\begin{align}
c_{r^2}&= \frac{\delta^4}{2 \Delta_c^5} \left( \mathfrak{C}_{1}\,p_{tc}^2+\mathfrak{C}_{2}\, p_{tc} L_{zc}+ \mathfrak{C}_{3}\, L_{zc}^2\right)\\
c_{t}&= \frac{\delta^2}{ \Delta_c^3} \left(\mathfrak{C}_{4}\, p_{tc} + \mathfrak{C}_{5}\, L_{zc}\right)\\
c_{t^2}&=-\frac{r_c }{2 \Delta_c}\left(a^2 (2 M+r_c)+r_c^3\right),\\
c_{z^2}&=\frac{r_c (r_c-2 M)}{2 \Delta_c},\\
c_{tz}&=-\frac{2 a M r_c}{\Delta_c},
\end{align}
where
\begin{align}
    \mathfrak{B}_{1}&= 2 M^2 \left(-a^4 (M-2 r_c)-2 a^2 r_c^3+M r_c^4\right)\\
\mathfrak{B}_{2}&= a M \left(a^4+a^2 \left(-4 M^2+8 M r_c-6 r_c^2\right)+r_c^4\right),\\
\mathfrak{B}_{3}&= a^2 (M-r_c) \left(a^2-2 M^2+2 M r_c-r_c^2\right),\\
\mathfrak{B}_{4}&= \left(a^2+r_c^2\right) \left(a^2 (M+r_c)+r_c^2 (r_c-3 M)\right),\\
\mathfrak{B}_{5}&= a M (a-r_c) (a+r_c),
\end{align}
and
\begin{widetext} 
\begin{equation} 
\begin{split}
\mathfrak{C}_{1}&= 4 M^2 \left(-a^6+2 a^4 \left(2 M^2-5 M r_c+5 r_c^2\right)-5 a^2 r_c^4+2 M r_c^5\right),\\
\mathfrak{C}_{2}&= 4 a M \left(a^4 (5 r_c-4 M)+2 a^2 \left(4 M^3-10 M^2 r_c+10 M r_c^2-5 r_c^3\right)+r_c^5\right),\\
\mathfrak{C}_{3}&=a^2  \left(a^4-2 a^2 \left(6 M^2-10 M r_c+5 r_c^2\right)+16 M^4-40 M^3 r_c+40 M^2 r_c^2-20 M r_c^3+5 r_c^4\right),\\
\mathfrak{C}_{4}&=\Delta_c^3+4 M^2 \left(a^4-3 a^2 r_c^2+2 M r_c^3\right),\\
\mathfrak{C}_{5}&=2 a  M \left(a^2 (2 M-3 r_c)+r_c^3\right).
\end{split}
\end{equation}
\end{widetext}

\section{Gravitational-wave fluxes}\label{sec:Teukolsky}

In this section, we describe our approach of calculating gravitational-wave fluxes using Teukolsky equation \cite{Teukolsky:1973ha} for geodesic orbits in the Kerr spacetime using action-angle formalism.

In the Teukolsky equation's framework, gravitational waves are treated as perturbations of the Kerr spacetime using Newmann-Penrose (NP) formalism. In this formalism, we calculate a perturbation of the NP scalar $\psi_4 = -C_{\alpha\beta\gamma\delta} n^\alpha \bar{m}^\beta n^\gamma \bar{m}^\delta$, where $C_{\alpha\beta\gamma\delta}$ is the Weyl tensor and $n^\mu$ and $\bar{m}^\mu$ are Kinnersley tetrad legs
\begin{align}
    n^\mu &= \left( r^2+a^2, -\Delta, 0, a \right)/(2\Sigma) \, , \\
    \bar{m}^\mu &= - \left( i a \sin\theta, 0, -1, i/\sin\theta \right)/(\sqrt{2}\zeta)
\end{align}
with $\zeta = r - i a \cos\theta$. This $\psi_4$ is governed by the Teukolsky equation \cite{Teukolsky:1973ha}, which we solve in frequency domain.

Because the radial and polar motion are recurrent,  the strain at infinity $h = h_+ - i h_{\times}$ can be written as a sum over discrete frequencies
\begin{equation}
\label{eq:strain}
    h = - \frac{2}{r} \sum_{lmkn} \frac{C^+_{lmkn}}{\omega_{mkn}^2} {}_{-2}S^{a\omega}_{lm}(\theta) e^{-i\omega_{mkn} u + i m \phi} \, ,
\end{equation}
where the frequencies are $\omega_{mkn} = m \Omega_\phi + k \Omega_\theta + n \Omega_r$, ${}_{-2}S^{a\omega}_{lm}(\theta)$ is the spin-weighted spheroidal harmonic function, $u = t-r^\ast$ is the retarded coordinate and the amplitudes can be expressed as two-dimensional integral \cite{Drasco06}
\begin{widetext}
\begin{multline} \label{eq:amplitudeint}
    C^+_{lmkn} = \frac{1}{2\pi\Upsilon} \int_0^{2\pi} {\rm d} \psi_r \int_0^{2\pi} {\rm d} \psi_\theta I^+_{lmkn}(r(\psi_r),\theta(\psi_\theta),u^r(\psi_r),u^\theta(\psi_\theta)) \times e^{i( k \psi_\theta + \omega_{mkn} \Delta t_\theta(\psi_\theta) - m \Delta \phi_\theta(\psi_\theta) )} \\ \times e^{i( n \psi_r + \omega_{mkn} \Delta t_r(\psi_r) - m \Delta \phi_r(\psi_r) )} \, ,
\end{multline}
\end{widetext}
where
\begin{equation}
    I^+_{lmkn} = \frac{\Sigma}{W} \sum_{i=0}^2 (-1)^i A_i \frac{d^i R^-_{lm\omega_{mkn}}}{dr^i} \, .
\end{equation}
$W = (R^+_{lm\omega}\partial_r R^-_{lm\omega}-R^-_{lm\omega}\partial_r R^+_{lm\omega})/\Delta$ is the invariant Wronskian and $R^{\mp}_{lm\omega}$ are the solutions of homogeneous radial Teukolsky equation satisfying ingoing (upgoing) boundary conditions (see, e.g., Eqs.~(92) in \cite{Pound22}). The functions $A_i$ are
\begin{align} 
    A_0 &= u_n^2 f^{(0)}_{nn} + u_n u_{\bar{m}} f^{(0)}_{n\bar{m}} + u_{\bar{m}}^2 f^{(0)}_{\bar{m}\bar{m}} \, , \\
    A_1 &= u_n u_{\bar{m}} f^{(1)}_{n\bar{m}} + u_{\bar{m}}^2 f^{(1)}_{\bar{m}\bar{m}} \, , \\
    A_2 &= u_{\bar{m}}^2 f^{(2)}_{\bar{m}\bar{m}}  \, ,
\end{align}
where $u_n$, $u_{\bar{m}}$ are projections of the four-velocity to the Kinnersley tetrad legs and the functions are
\begin{align}
    f^{(0)}_{nn} &= -\frac{2\zeta^2}{\Delta^2} \left( \mathcal{L}^\dag_1 \mathcal{L}^\dag_2 -2ia\zeta^{-1} \sin\theta \mathcal{L}^\dag_2 \right)S \, , \\
    f^{(0)}_{n\bar{m}} &= \frac{2\sqrt{2}\zeta^2}{\bar{\zeta} \Delta} \bigg( \left( \frac{i K}{\Delta} + \zeta^{-1} + \bar{\zeta}^{-1} \right) \mathcal{L}^\dag_2 \nonumber\\ &\hphantom{=} - a \sin\theta \frac{K}{\Delta} \left( \bar{\zeta}^{-1} - \zeta^{-1} \right) \bigg)S \, , \\
    f^{(1)}_{n\bar{m}} &= \frac{2\sqrt{2}\zeta^2}{\bar{\zeta} \Delta} \left( \mathcal{L}^\dag_2 + i a \sin\theta \left( \bar{\zeta}^{-1} - \zeta^{-1} \right) \right)S \, , \\
    f^{(0)}_{\bar{m}\bar{m}} &= \frac{\zeta^2}{\bar{\zeta}^2} \left( i \partial_r \left( \frac{K}{\Delta} \right) - 2 i \zeta^{-1} \frac{K}{\Delta} + \left( \frac{K}{\Delta} \right)^2 \right)S \, , \\
    f^{(1)}_{\bar{m}\bar{m}} &= - \frac{2 \zeta^2}{\bar{\zeta}^2} \left( \zeta^{-1} + i \frac{K}{\Delta} \right)S \, , \\
    f^{(2)}_{\bar{m}\bar{m}} &= - \frac{\zeta^2}{\bar{\zeta}^2} S  \, ,
\end{align}
where $S = {}_{-2}S^{a\omega_{mkn}}_{lm}(\theta)$, $K=(r^2+a^2)\omega - am$ and
\begin{equation}
    \mathcal{L}^\dag_s = \frac{\partial}{\partial\theta} -\frac{m}{\sin\theta} + a \omega \sin\theta + s\cot\theta \, .
\end{equation}

Thanks to the flux-balance laws, the rate of loss in $E$ and $L_z$ is equal to the GW fluxes of energy and angular momentum to infinity and through the horizon. Similar law holds for the Carter constant $Q$. Explicit expressions can be found e.g. in Eqs. (3.26)-(3.32) of \cite{Hughes21}.

We have developed a new code in Mathematica which solves the inhomogeneous Teukolsky equation in the frequency domain. The transformation relations between BL coordinates and the phases $r(\psi_r)$ and $\theta(\psi_\theta)$ as well as the oscillating parts $\Delta t_{r,\theta}$ and $\Delta \phi_{r,\theta}$ are provided in the \textit{CPKerrGeodesics} package. For the integration in Eq.~\eqref{eq:amplitudeint}, the midpoint rule was employed since the integrand is periodic in $\psi_r$ and $\psi_\theta$ and, therefore, it has exponential convergence. To find the homogeneous solutions of the radial and angular Teukolsky equation $R^\pm_{lm\omega}$ and ${}_{-2}S^{a\omega}_{lm}$ we used the Black Hole Perturbation Toolkit \cite{BHPT}. 

\subsection{Flux Grid}

\begin{figure}
    \centering
      \includegraphics[width=0.483\textwidth]{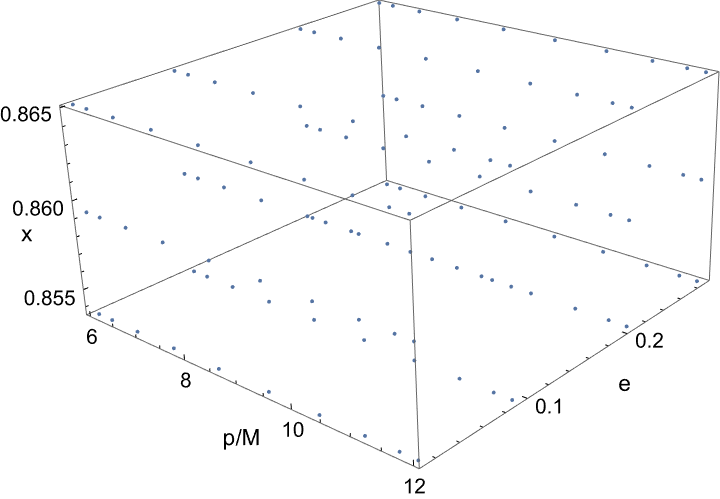}
    \caption{Grid for interpolating the fluxes in the $(p,e,x)$ space,  $p\in (6,12)$, $e\in (0,0.3)$ and $x \in (\cos(31.5{}^{\circ}),\cos(30{}^{\circ})) $. } \label{fig:grid}
\end{figure}

Let us now briefly comment on the calculation of the total fluxes of the constants of motion. The averaged rate of change of a constant of motion $C$ is given by the corresponding fluxes to infinity and horizon
\begin{align}
\Big\langle\frac{\mathrm{d}C}{\mathrm{d}t}\Big\rangle= -\Big(F^{\infty}+F^{H} \Big).
\end{align}
The fluxes  can be in the same fashion as the strain~\eqref{eq:strain}  expressed as sum of individual modes
\begin{align}
F=\displaystyle\sum_{l,m,k,n} F_{l m k n}.
\end{align}
Each of the $F_{lmkn}$ is determined by the amplitudes $C^\pm_{lmkn}$ discussed above.
During our calculation, we use the symmetry  $F_{lmkn}=F_{l-m-k-n}$. We, thus, sum only the modes with $\omega_{mkn}>0$ and then multiply the result by $2$.

\begin{figure*}[ht]
\begin{center}
 {\subfloat{\includegraphics[width=0.48\textwidth]{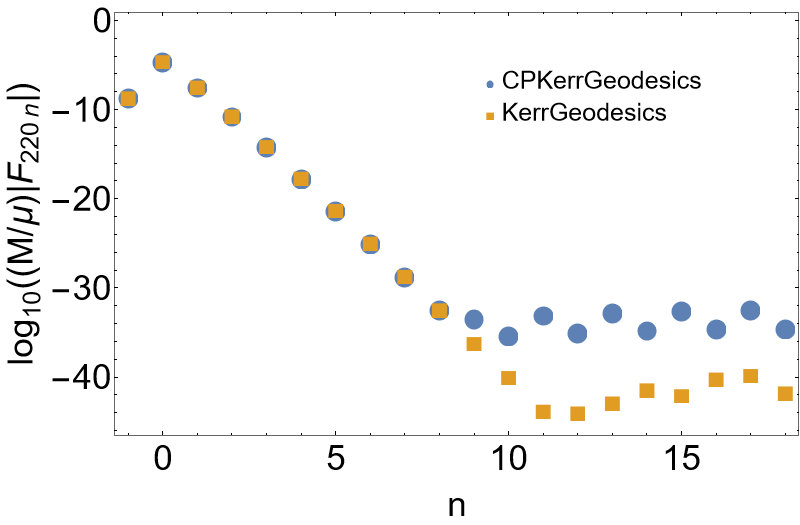}}
{  \subfloat{\includegraphics[width=0.48\textwidth]{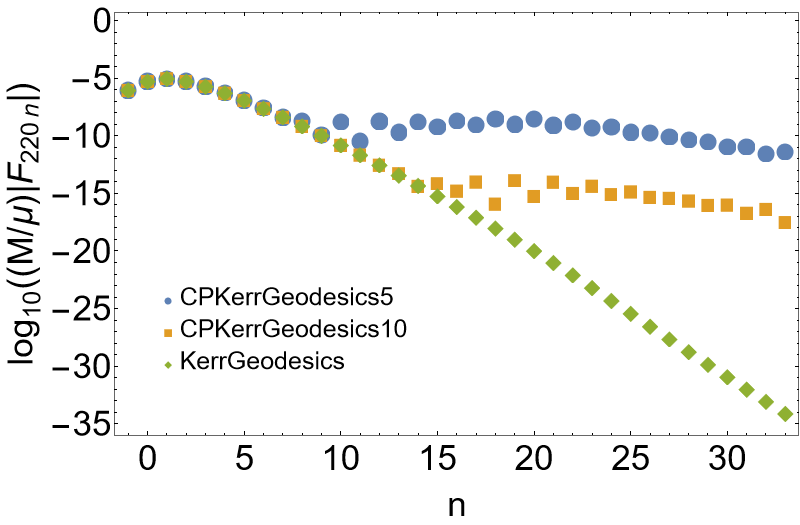}}}
}   
\end{center}
   \caption{Logarithmic plots  comparing the convergence of partial fluxes  $F_{220n}$ (more specifically, the energy fluxes to infinity) for the KerrGeodesics package with our CPKerrGeodesics package. The orbital parameters are $a=0.5 M$, $p=10M$, $\iota_0= \pi/6$ with eccentricities $e=0.01$ (left) and  $e=0.3$ (right).}
        \label{fig:convergence}
\end{figure*}

Our summation algorithm starts from the dominant  $F_{220n}$  modes where we start by calculating the contributions for growing $n$. We stop this procedure at some $n_{0}$  once $F_{220n}$ stops converging to zero and starts oscillating as shown in Fig.~\ref{fig:convergence}. This happens for higher values of $n$ and the exact value of $n_{0}$ depends on the eccentricity; for higher eccentricities, we have to sum more modes so $n_{0}$ is larger for larger $e$. This error is introduced by our perturbative approximation to geodesics. All the modes with contribution smaller than  $F^{\infty}_{220n_{0}}+F^{H}_{220n_{0}}$ are omitted in the subsequent summation. We then sum over $k$  
\begin{align*}
   F_{22}=\displaystyle\sum^{3}_{k=-3} F_{22k },\enspace F_{22k }=\displaystyle\sum_{n}F_{22kn}  
\end{align*}
For the $m=2$ mode, we sum over $l$ from $2$ to $10$. 

To shorten the time of the calculation for the other modes, it is useful to estimate the maximum value $\displaystyle\max_{k} F_{l m k }=F_{l m k_{\rm max}}$ with $F_{lmk}=\displaystyle\sum_{n}F_{lmkn}$, when summing over $k$ for fixed $m$ and $l$.   $F_{l m k }$ for fixed $m$ and $l$ seems to have a maximum when $k_{\rm max}=l-m$. In our summation the index $k$, then goes from $k_{\rm max}-3$ to $k_{\rm max}+3$. Having settled the summation over $k$, we then sum over $m$ from $-3$  to $6$ with $l$ going from  $\max(2,\vert m\vert)$ to $10$. Regardless of our summation scheme, it is important to stress again that the deviation from the correct values of fluxes is dominantly caused by our approximation of the geodesics and not by neglecting higher modes.

Even if this was already discussed, it is useful to compare the partial fluxes $F_{lmkn}$ computed using our perturbatively derived geodesics to those calculated from the exact geodesics using the same Teukolsky solver just like we did in Sec.~\ref{sec:inspiral} for the total fluxes. If we fix the indices $l$, $m$, $k$, the expected behaviour would be $F_{lmkn}\rightarrow 0$ as $n\rightarrow \infty$. When looking at the logarithmic plots~\ref{fig:convergence} depicting the dominant modes  $F_{220n}$, we can indeed see that the partial fluxes decrease with $n$, but due to numerical errors even the fluxes calculated using the KerrGeodesics package eventually start oscillating around a small but non-zero value. When close to the spherical orbits we can see that this happens for small $n$, but this is not an issue as the higher modes barely contribute to the sum (for an exact spherical orbit there is no sum over $n$). For eccentric orbits more $n$-modes need to be summed and here there is a clear difference between the exact geodesics and our approximation.
 As the oscillations start sooner (for $n=n_0$) when using the approximation, we are forced  to neglect all modes with $n>n_0$ which is the primary source of deviation  from the correct values of the total fluxes. By adding more generating functions, we can push  $n_0$ to higher values which is illustrated in Fig.~\ref{fig:convergence} (5 versus 10 generating functions). For higher modes the integrand in Eq.~\eqref{eq:amplitudeint} has more and more oscillations which tend to cancel each other, leaving us with a very small resulting number, which means that even a small deviation in the integrand can create a large  error in the  Teukolsky amplitude and consequently in the partial fluxes.

To calculate the fluxes we used a grid and Chebyshev interpolation like in  \cite{Skoupy22}. In particular, we created a $10\times4\times3$ grid in the $(p,e,x)$ space for $a=0.5 M$ (see Fig.~\ref{fig:grid}) with the interpolation points located at the Chebyshev nodes. Note that the grid  is simpler than in \cite{Skoupy22} since we do not come very close to the separatrix. The relative  error  of the  Chebyshev interpolation is  of the order $10^{-3}$, which is similar to the relative errors of the fluxes themselves. Since there are analytical formulas $C_i=C_i(p,e,x)$ linking the constants of motion to the orbital parameters, it is possible to differentiate them and transform the fluxes of $C_i$ to fluxes of $(p,e,x)$, this is in detail described in \cite{Hughes21}.

\end{document}